\documentclass[final,5p,times,twocolumn]{elsarticle}

\usepackage{hyperref}
\usepackage{mathrsfs}
\hypersetup{
    colorlinks=true,       
    linkcolor=blue,          
    citecolor=blue,        
    filecolor=blue,      
    urlcolor=blue           
}

\usepackage{graphicx}
\usepackage{amsmath}
\usepackage{amssymb}
\usepackage{braket}
\usepackage{color}

\graphicspath{{./}{./figs/}}

\newcounter{bla}

\journal{Computer Physics Communications}

\begin{document}

\begin{frontmatter}



\title{PyLCP: A python package for computing laser cooling physics}


\author[a]{Stephen Eckel\corref{author}}
\author[a]{Daniel S. Barker}
\author[b]{Eric B. Norrgard}
\author[a]{Julia Scherschligt}

\cortext[author] {Corresponding author.\\\textit{E-mail address:} stephen.eckel@nist.gov}
\address[a]{Sensor Science Division, National Institute of Standards and Technology, Gaithersburg, MD 20899, USA}
\address[b]{Joint Quantum Institute, National Institute of Standards and Technology
and University of Maryland, Gaithersburg, Maryland 20899, USA}

\begin{abstract}
We present a python object-oriented computer program for simulating various aspects of laser cooling physics.
Our software is designed to be both easy to use and adaptable, allowing the user to specify the level structure, magnetic field profile, or the laser beams' geometry, detuning, and intensity.
The program contains three levels of approximation for the motion of the atom, applicable in different regimes offering cross checks for calculations and computational efficiency depending on the physical situation.
We test the software by reproducing well-known phenomena, such as damped Rabi flopping, electromagnetically induced transparency, stimulated Raman adiabatic passage, and optical molasses.
We also use our software package to quantitatively simulate recoil-limited magneto-optical traps, like those formed on the narrow $^1$S$_0\rightarrow ^3$P$_1$ transition in $^{88}$Sr and $^{87}$Sr.
\end{abstract}


\end{frontmatter}


{\bf PROGRAM SUMMARY}

\begin{small}
\noindent
{\em Program Title: PyLCP}                                          \\
{\em Licensing provisions(please choose one): NIST}                                   \\
{\em Programming language: Python}                                   \\

{\em Nature of problem:} Accurate simulation of laser cooling physics presents a non-linear problem that is highly dependent on atomic structure and a host of control parameters including laser beam geometry, detuning, intensity, transverse mode, polarization and magnetic field profile.  Exploration of the parameter space can be difficult, as most laser-cooling simulations to date have been coded only for specific Hamiltonians and/or laser and magnetic fields.\\

{\em Solution method:}
An object-oriented python package that allows for easy modification of the control parameters and internal structure of the atom followed by automated construction of the governing equations to simulate the particular experimental interest.\\

\end{small}

\section{Introduction}
\label{sec:intro}
Laser cooling is ubiquitous in modern atomic physics.
Cooling, trapping, and manipulating atoms has led to advances in clocks~\cite{Ludlow2015, Campbell2017, Elvin2019}, inertial sensors~\cite{Cronin2009, Becker2018, Xu2019}, magnetometers~\cite{Kitching2018, Cohen2019}, emerging quantum technologies~\cite{Pezze2018, Bruzewicz2019}, and tests of fundamental symmetries~\cite{Safronova2018, hutzler2020, Norrgard2019}.
On its surface, laser cooling appears straightforward, with well-known models that qualitatively describe its essential features, like magneto-optical trapping~\cite{Foot2005, Metcalf1999} and Doppler~\cite{Lett1989} and sub-Doppler cooling~\cite{Dalibard1989}.
As progress is made toward deploying laser-cooled atoms in photonically integrated packages~\cite{Rushton2014}, modelling is required in order to optimize design parameters and understand engineering tolerances.
At the same time, progress is being made on theoretically understanding new techniques of laser cooling~\cite{Bartolotta2018}.
For both of these applications, a suite of simple-to-use software capable of simulating the full description of laser cooling, including complicated level structures and complex trapping geometries, is still lacking.

For example, consider a $^{23}$Na atom in a standard, six-beam magneto-optical trap (MOT)~\cite{Raab1987}.
The trap consists of a spherical quadrupole magnetic field and six independent laser beams with two frequency components that drive transitions between 28 different Zeeman states.
The dynamics of a $^{23}$Na atom in a MOT are well-described by the optical Bloch equations (OBE), a comprehensive framework that models classical fields coupled to non-interacting atoms and includes internal state coherences.
The OBEs constitute a total of $28^2=784$ coupled, first-order, time and position dependent differential equations for the atom's internal states (plus an additional 6 differential equations to account for its classical motion).
Automatically generating and efficiently solving the OBEs for an arbitrary combination of atomic species, laser fields and magnetic fields is a necessity for theoretically quantifying the quality of different kinds of traps.

We introduce an open-source, python-based program that computes the movement of atoms or molecules with complex level structures in arbitrary optical (laser) and magnetic fields.
The {\tt pylcp} package allows multiple levels of approximation from the complete OBEs through to a simple heuristic model.
Like other quantum dynamics packages~\cite{Johansson2012, Johansson2013, ADM}, {\tt pylcp} can solve the optical Bloch equations, but it focuses on simulating the laser cooling of atoms and molecules.
Importantly, for the user's given laser geometry, atomic level structure, and magnetic field configuration, our code automatically generates the governing equations for the atom.
We leverage existing python packages to both integrate the resulting equations and analyze their output.
Our program is object-oriented, allowing simple extensions to include additional types of laser beams and magnetic fields, along with different atomic and molecular Hamiltonians.
The source code and documentation for {\tt pylcp} are freely available online~\cite{pylcp, pylcpdocs}, with stable releases distributed via the python package index (PyPI).

The remainder of this paper is organized as follows.
In Sec.~\ref{sec:governing_equations}, we present the governing equations included in the {\tt pylcp} package.
We start with the OBEs, then present approximations that result in the rate equations and further approximations that result in a heuristic model.
In each derivation, we focus on the elements that are important for its efficient programming.
In Sec.~\ref{sec:examples}, we present several tests of the code against textbook examples and find quantitative agreement with well-known results.
We also use our software package to calculate properties of more exotic MOTs, including those formed on the narrow $^1$S$_0\rightarrow ^3$P$_1$ transition of $^{88}$Sr and $^{87}$Sr.

\section{The governing equations}
\label{sec:governing_equations}
\subsection{The optical Bloch equations}
Our derivation for the OBEs follows those in Refs.~\cite{Gordon1980, Ungar1989, Devlin2018}, which can be consulted for further detail.
We consider the generic problem of coupling $N$ quantum states together in arbitrary optical and magnetic fields.
We group the states into manifolds: a collection of states that are degenerate or nearly degenerate, e.g., the hyperfine states of the $^2S_{1/2}$ state of an alkali atom or the ro-vibrational states of a molecule.
We denote the $i$th state and its manifold index $n$ by  $\ket{i, n}$.
The manifolds are useful both for defining appropriate rotating frames and for applying the rotating wave approximation.

The full Hamiltonian is given by
\begin{equation}
    \label{eq:obe:generic_hamiltonian}
    \hat{H} = \hat{H}_\text{atom} + \hat{H}_\text{field} -
    \hat{\mathbf{d}}\cdot\hat{\mathbf{E}} -
    \hat{\boldsymbol{\mu}}\cdot\hat{\mathbf{B}}.
\end{equation}
The atomic operators $\hat{\mathbf{d}}$ and $\hat{\boldsymbol{\mu}}$ are the electric and magnetic dipole operators, respectively\footnote{Throughout this paper, bold face denotes vector quantities.  Hats usually denote quantum mechanical operators, with the exception of $\hat{x}$, $\hat{y}$, and $\hat{z}$.  In this case, the hat denotes the relevant Cartesian unit vector.}.
The field component of the Hamiltonian is given by
\begin{equation}
    \label{eq:obe:field}
    \hat{H}_\text{field} = \int \left(\frac{\epsilon_0 \hat{\mathbf{E}}}{2} +
    \frac{\hat{\mathbf{B}}}{2\mu_0}\right)\ dV,
\end{equation}
where $\hat{\mathbf{E}}$ is the electric field operator, $\hat{\mathbf{B}}$ is the magnetic field operator, \(\epsilon_0\) is the vacuum permittivity, and \(\mu_0\) is the vacuum permeability.
The atomic Hamiltonian is
\begin{equation}
    \hat{H}_\text{atom} = \frac{P^2}{2M} + \hat{H}_\text{int},
\end{equation}
where $\hat{H}_\text{int}$ describes the atom's internal structure, \(\mathbf{P}\) is its momentum, and \(M\) is its mass.
In general, \(\hat{H}_\text{int}\) has the form
\begin{equation}
    \label{eq:H_internal}
	\hat{H}_\text{int} = \sum_{ij} \hbar \omega_n\ket{i,n}\Bra{i, n} + (H_n)_{ij}\ket{i,n}\Bra{j, n},
\end{equation}
where $(H_n)_{ij} = \Braket{i,n|\hat{H}^n_{\text{int}}|j,n}$ and $\omega_n$ is the offset frequency of the $n$th manifold.
$H_n$ is a matrix that need not be diagonal.
Manifolds are connected only through the $\hat{\mathbf{d}}\cdot\hat{\mathbf{E}}$ component of the Hamiltonian; $\hat{H}^n_\text{int}$ and $\hat{\boldsymbol{\mu}}\cdot\hat{\mathbf{B}}$ only act on the subspace of each manifold.

Our goal is to find the evolution of the expectation values of the density operators $\hat{\rho}_{ij} = \ket{i}\Bra{j}$.
(We suppress the manifold index when it is not relevant.)
In the Heisenburg picture, an operator $\hat{O}$ evolves as
\begin{equation}
	\label{eq:heisenburg_evolution}
	\frac{\partial \hat{O}}{\partial t}  = \frac{i}{\hbar}[\hat{H}, \hat{O}].
\end{equation}
If the fields were treated classically, this equation would have to be amended in order to take into account decays.
Instead, if we keep the fields quantized, derive the equations of motion, and then apply appropriate radiation reaction approximations, we can derive the full OBEs with decay included from \eqref{eq:heisenburg_evolution}.
The magnetic field $\mathbf{B}$ is assumed to be a classical field; we will not consider quantizing it.
We must pay special focus to the electric field $\mathbf{E}$, however, for it both shifts the internal Hamiltonian when transformed into the necessary rotating frame(s) and creates the necessary decay channels.

For the electric field, $\mathbf{E}$ could be comprised of multiple modes.
We group those modes by the $n\rightarrow m$ transitions they drive between manifolds $n$ and $m$.
For each manifold pair, we extract a carrier frequency $\omega_{n\rightarrow m}$.
Thus, the electric field of the mode driving $n\rightarrow m$ is
\begin{equation}
	\hat{\mathbf{E}} = \hat{\mathbf{E}}_{n\rightarrow m} e^{-i \omega_{n\rightarrow m} t} + \hat{\mathbf{E}}_{n\rightarrow m}^\dagger e^{i \omega_{n\rightarrow m} t}.
\end{equation}
Here, $\hat{\mathbf{E}}_{n\rightarrow m}$ represents a destruction operator of the mode $n\rightarrow m$.
We note that $\hat{\mathbf{E}}_{n\rightarrow m}$ can itself be time-dependent and include additional sub-modes with frequencies $\omega_p \ll \omega_{n\rightarrow m}$, but these are assumed to be captured by the time dependence of $\hat{\mathbf{E}}_{n\rightarrow m}$.

We similarly expand the dipole operator
\begin{equation}
	\hat{\mathbf{d}} = \sum_{ijnm}\mathbf{d}^{nm}_{ij}\ket{n, i}\Bra{j, m} + \mathbf{d}^{*nm}_{ji}\ket{j,m}\Bra{i,n}.
\end{equation}
Here, $\mathbf{d}^{nm}_{ij} = \Braket{i,n|\hat{\mathbf{d}}^{nm}|j, m}$ and $\hat{\mathbf{d}}^{nm}$ is the dipole matrix operator between the two manifolds $n$ and $m$ with $m$ being larger in energy than $n$.
We further define the complex conjugate $\mathbf{d}^{*nm}_{ji} = \Braket{j,m|\hat{\mathbf{d}}^{nm}|i, n}$, where we have reversed only the specific indices of the state\footnote{We do not reverse $nm$ ordering as that merely denotes what manifolds $\mathbf{d}$ connect.}.
In general, $\mathbf{d}^{nm}$ is a vector of matrices.
We choose to express this vector in the spherical basis, with vector elements $d^{nm}_q$.
The individual matrix elements $(d^{nm}_q)_{ij}$ are dependent on reduced matrix elements and Clebsch-Gordan coefficients that determine the transitions between manifolds $n$ and $m$.
We will not consider any specific form of $\mathbf{d}^{nm}_{ij}$, but instead focus on deriving the OBEs for any generic $\mathbf{d}^{nm}$.

To determine the decays, we must apply a radiation reaction approximation.
Classically, the radiation reaction field is
\begin{equation}
	\mathbf{E}_{RR} = \frac{1}{6\pi\epsilon_0 c^3}\frac{d^3\mathbf{d}}{dt^3}.
\end{equation}
The dipole moment $\mathbf{d}$ will oscillate with all frequency components contained in the drive.  Thus, for each frequency mode, we must take $d$ to have an $e^{-i \omega_{n\rightarrow m} t}$ oscillation and the total electric field operator becomes
\begin{equation}
	\hat{\mathbf{E}}_{n\rightarrow m} = \hat{\mathbf{E}}_{0,n\rightarrow m} + \frac{i \omega_{n\rightarrow m}^3}{6\pi\epsilon_0c ^3}e^{-i\delta_{n\rightarrow m, p}t} \mathbf{d}^{nm}_{ij}\ket{i}\Bra{j},
\end{equation}
where $\mathbf{E}_{0,n\rightarrow m}$ is the operator on the classical field.
The two density operators each contribute their preferred rotation, yielding the total oscillation of $\delta_{n\rightarrow m} = \omega_{n\rightarrow m}-(\omega_{m}-\omega_{n})$.
We then note that for each manifold, any additional frequency component $\omega_p\ll \omega_{n\rightarrow m}$, so the wavevector is assumed to be the same for the entire manifold, making
\begin{equation}
	\label{eq:total_e_field_operator}
	\hat{\mathbf{E}}_{n\rightarrow m} = \hat{\mathbf{E}}_{0,n\rightarrow m} + \frac{i k_{n\rightarrow m}^3}{6\pi\epsilon_0 }e^{-i\delta_{n\rightarrow m, p}t} \mathbf{d}^{nm}_{ij}\ket{i}\Bra{j},
\end{equation}
where $k_{n\rightarrow m} = \omega_{n\rightarrow m}/c$.
It is now interesting to note the commutation relationships.
Clearly the $\rho_{ij}$ operators must commute with all $\hat{\mathbf{E}}_{n\rightarrow m}$, as they are different physical observables.
On the other hand, given that $\rho_{ij}$ does not commute with the second term in (\ref{eq:total_e_field_operator}), it must also not commute with the classical field operator $\hat{\mathbf{E}}_{0,n\rightarrow m}$.

For each manifold, we will assume that all states in that manifold rotate at a preferred frequency $\ket{i,n} \rightarrow e^{i \omega^R_n t}\ket{i, n}$.
We choose the $\omega^R_n$ such that their differences $\omega^R_m-\omega^R_n \approx \omega_m-\omega_n \approx \omega_{n\rightarrow m}$ for all combinations of $n$ and $m$.
This choice places each manifold into an appropriate rotating frame.
Under this transformation of the state vectors, the internal Hamiltonian becomes
\begin{equation}
	\hat{H}_\text{int} = -\hbar \delta^H_n \ket{i,n}\Bra{i, n} +  H^n_{\text{int}, ij}\ket{i,n}\Bra{j, n},
\end{equation}
where we define $\delta^H_n = \omega^R_n-\omega_n$, incorporating the shift into the rotating frame into the internal Hamiltonian.
We also define $\delta^L_{n\rightarrow m} = \omega_{n\rightarrow m} - (\omega^R_m - \omega^R_n)$.  Making the rotating wave approximation (neglecting terms oscillating at optical frequencies), we find, keeping only energy conserving terms,
\begin{eqnarray}
    \hat{\mathbf{d}}\cdot\hat{\mathbf{E}} & = & \left(\mathbf{d}^{nm}_{ij}\cdot\hat{\mathbf{E}}^\dagger_{n\rightarrow m} e^{-i\delta^L_{n\rightarrow m}  t} \right)  \ket{i,n}\Bra{j,m} +  \nonumber \\
	& & ~~~~~~ \ket{j,m}\Bra{i,n} \left((\mathbf{d}^{*nm}_{ji}) \cdot \hat{\mathbf{E}}_{n\rightarrow m}e^{i\delta^L_{n\rightarrow m}t}\right).\label{eq:d_dot_E}
\end{eqnarray}
Note that because $\mathbf{E}$ commutes with $\rho_{ij}$, we can place the operators in any order.
We have chosen normal order: the first operator to apply to the wavefunction is the destruction operator of either the atom or the field and the second operator is creation operator.
This operator ordering is required for the radiation reaction approximation to produce the correct decay rate $\Gamma$~\cite{Dalibard1982, Milonni1984}.

It is instructive to consider two examples of this construction of the rotating frame(s).
Consider first a standard two level system, with indices $i=g$ and $j=e$ and energies $\omega_g=0$ and $\omega_e$, being driven by a single electric field with frequency $\omega$.
Assume the states rotate at preferred frequencies defined by $\omega^R_e$ and $\omega^R_g=0$.
Then the detuning on the Hamiltonian, $\delta^{H}_e = \omega^R_e - \omega_e$ and $\delta^L_{g\rightarrow e} = \omega - \omega^R_e$.
The total detuning of the laser from the excited state is then given by $\delta = \delta^L_{g\rightarrow e}+\delta^H_e = \omega-\omega_e$.
In this way, we can split the detuning between lasers and Hamiltonian in whichever way yields best computational efficiency for the problem at hand.

Next, consider a three manifold $\Lambda$-system with a single state in each manifold.
Let us label the manifolds as $g$, $r$ and $e$ in order of overall energy, and drop the unnecessary substate subscripts.
We address this system with two lasers, one tuned closely to $g\rightarrow e$ with frequency $\omega_{g\rightarrow e}$ and the other tuned closely to $r\rightarrow e$ with frequency $\omega_{r\rightarrow e}$.
We now choose $\omega^R_g = \omega_g = 0$, and the relevant detunings are then
\begin{eqnarray}
	\delta^H_g & = & 0 \\
	\delta^H_e & = & \omega^R_e - \omega_e \\
	\delta^H_r & = & \omega^R_r - \omega_r \\
	\delta^L_{g\rightarrow e} & = & \omega_{g\rightarrow e} - \omega^R_e \\
	\delta^L_{r\rightarrow e} & = & \omega_{r\rightarrow e} - (\omega^R_e-\omega^R_r).
\end{eqnarray}
By choosing $\omega^R_e = \omega_{g\rightarrow e}$ and $\omega^R_r = \omega_{g\rightarrow e} - \omega_{r\rightarrow e}$, one recovers the textbook example of the three level system with detunings appearing entirely on the Hamiltonian.
As with the two level system above, one can split the detunings between lasers and Hamiltonian in whichever way yields best computational efficiency for the problem at hand.

Ref.~\cite{Devlin2018} uses a choice of rotating frame that is equivalent to having one rotating frame per set of degenerate levels at zero field.
This choice is conceptually easier to understand and potentially computationally advantageous, as it eliminates many high-frequency oscillations in the Hamiltonian.
However, its construction has disadvantages as well, most notably not being able to include non-linear Zeeman couplings between different zero-field degenerate states.
Nevertheless, between {\tt pylcp} and the construction of Ref.~\cite{Devlin2018}, the largest difference appears to be the inclusion of oscillatory terms in the decay between excited states and ground states.
Those terms are neglected here, as they are related to the frequency differences between different rotating frames, which, in our construction, is assumed to be large.

Before applying the radiation reaction approximation, we must first find the equations of motion.
Inserting \eqref{eq:d_dot_E} into \eqref{eq:heisenburg_evolution}, and using $\hat{\rho}_{ij}\hat{\rho}_{kl} = \hat{\rho}_{il}\delta_{jk}$, where $\delta_{ij}$ is the Kronicker delta function,
\begin{eqnarray}
	\hbar\frac{\partial \hat{\rho}_{ij}}{\partial t}
	& = & -i\left(\mathbf{d}^{nm}_{ki}\cdot\hat{\mathbf{E}}^\dagger_{n\rightarrow m} e^{-i\delta^L_{n\rightarrow m}  t} \right)  \ket{k}\Bra{j} +  \nonumber \\
	& & -i \ket{k}\Bra{j} \left(\mathbf{d}^{*nm}_{ki} \cdot \hat{\mathbf{E}}_{n\rightarrow m}e^{i\delta^L_{n\rightarrow m}t}\right) \nonumber \\
	& & +i \left(\mathbf{d}^{nm}_{jk}\cdot\hat{\mathbf{E}}^\dagger_{n\rightarrow m} e^{-i\delta^L_{n\rightarrow m}  t} \right)\ket{i}\Bra{k}  \nonumber \\
	& & +i \ket{i}\Bra{k} \left(\mathbf{d}^{*nm}_{jk} \cdot \hat{\mathbf{E}}_{n\rightarrow m}e^{i\delta^L_{n\rightarrow m}t}\right). \label{eq:E_coherent_ev}
\end{eqnarray}
Once again, we have maintained normal operator order.
Taking the expectation value of \eqref{eq:E_coherent_ev} and focusing on the real part of evolution results in
\begin{equation}
\label{eq:decay}
\begin{aligned}
	\hbar\,\text{Re}\left(\frac{\partial \rho_{ij}}{\partial t}\right) =  
	\frac{k_{n\rightarrow m}^3}{6\pi\epsilon_0 }&\left[
	-\mathbf{d}_{ki}\cdot \mathbf{d}^*_{lk}\rho_{lj}
	+ \mathbf{d}_{ki}^* \cdot \mathbf{d}_{jl}\rho_{kl} \right. \\
	& ~\left. + \mathbf{d}_{jk}\cdot \mathbf{d}^*_{li}\rho_{lk}
	- \mathbf{d}_{jk}^* \cdot \mathbf{d}_{kl}\rho_{il} \right],
\end{aligned}
\end{equation}
where $\rho_{ij} = \langle \hat{\rho}_{ij} \rangle$.  This equation defines the decay in terms of the $\mathbf{d}^{nm}$ matrices.
{\tt pylcp} represents the decay evolution as a matrix equation, $\dot{\boldsymbol{\rho}} = (1/\hbar) \boldsymbol{\Gamma} \cdot \boldsymbol{\rho}$, where $\boldsymbol{\rho}$ is a flattened vector of the expectation values of all $\rho_{ij}$.
The contribution of each dipole operator connecting two manifolds is calculated separately; namely, $\boldsymbol{\Gamma} = \sum_{m,n<m} \boldsymbol{\Gamma}^{m\rightarrow n}$ where $\boldsymbol{\Gamma}^{m\rightarrow n}$ is calculated using \eqref{eq:decay} from $\mathbf{d}^{nm}$.

The remaining part of the $\hat{\mathbf{d}}\cdot\hat{\mathbf{E}}$ operator is the imaginary component, which describes the coherences between the quantum states of different manifolds.
The imaginary component is proportional to $\hat{\mathbf{E}}_{0,n\rightarrow m}$.
We use the result $\hat{\mathbf{E}}_{0,n\rightarrow m}\ket{\mathbf{E}(\mathbf{r})} = \mathbf{E}_{0,n\rightarrow m}(\mathbf{r},t) \ket{\mathbf{E}_{0,n\rightarrow m}(\mathbf{r})}$, where $\mathbf{E}_{0,n\rightarrow m}(\mathbf{r},t)$ is the classical field.
When computing the expectation value of the classical field with this identity, the only operators that remain in \eqref{eq:E_coherent_ev} are the density operators.
Therefore, we can compute the evolution due to the classical electric field in the same manner as that of the classical magnetic field.

To calculate imaginary part of the evolution due to all fields and $\hat{H}_{int}$, we insert $\hat{O}=\hat{\rho}$ and the full $\hat{H}$ into~\eqref{eq:heisenburg_evolution}, using the classical field values for $\mathbf{B}$ and $\mathbf{E}$.
This insertion naturally includes the imaginary part of~\eqref{eq:E_coherent_ev}.
Computing the expectation value, \eqref{eq:heisenburg_evolution} becomes a matrix equation, where the individual elements are given by
\begin{equation}
    \label{eq:im_evolution_H}
	\hbar\,\text{Im}\left(\frac{\partial \rho_{ij}}{\partial t}\right) = -\sum_k\left(\rho_{ik}H_{kj} - H_{jk} \rho_{ki}\right),
\end{equation}
where $H_{ij} = \langle \hat{H}_{ij} \rangle$ is the expectation value of the Hamiltonian.
Because the classical fields (i.e, the expectation value of the classical field operators) commute with the atomic density operators, we split the evolution between $H_\text{int}$, $\mathbf{\mu}_n$, and $\mathbf{d}^{nm}$, calculating the commutator for each separately.
As with $\boldsymbol{\mathbf{\Gamma}}$, {\tt pylcp} represents \eqref{eq:im_evolution_H} as a matrix equation $\dot{\boldsymbol{\rho}} = -(i/\hbar) \boldsymbol{\mathcal{H}_\text{int}}\cdot \boldsymbol{\rho}$, where the corresponding elements of the matrix $\boldsymbol{\mathcal{H}_\text{int}}$ are calculated from \eqref{eq:im_evolution_H} using the substitution $\hat{H} = \hat{H}_\text{int}$.
Likewise, matrices $\boldsymbol{\cal D}^{nm}_q$, $\boldsymbol{\cal D}^{*,nm}_q$, and $\boldsymbol{\cal M}_{n,q}$ are calculated for the evolution due to $d^{nm}_q$, $d^{*nm}_q$, and $\mu_{n,q}$, respectively.

The full evolution is then given by
\begin{eqnarray}
    \dot{\boldsymbol{\rho}} & = & \frac{1}{\hbar}\left[\boldsymbol{\Gamma}-i\left(\boldsymbol{\mathcal{H}}_\text{int} - \sum_{q,n,m}(-1)^q [\boldsymbol{\cal D}^{nm}_qE^*_{0, n\rightarrow m, -q} \right.\right. \nonumber\\
    & & \left.\left.
     ~~~~~~~~~~~~ - \boldsymbol{\cal D}^{*nm}_q E_{0, n\rightarrow m,-q}  - \boldsymbol{\cal M}_{n,q}B_{-q}]\right)\right]\cdot \boldsymbol{\rho}.
\end{eqnarray}
Because commutators are computationally intensive, the matrices $\boldsymbol{\Gamma}$,  $\boldsymbol{H}_\text{int}$, $ \boldsymbol{\cal D}^{nm}_q$, $\boldsymbol{\cal D}^{*,nm}_q$, and $\boldsymbol{\cal M}_{n,q}$  are only calculated once when the OBEs are initially generated and therefore cannot be time dependent.
The OBEs are symmetric under population exchange, i.e., $\rho_{ij} = \rho_{ji}^*$.
Optionally, {\tt pylcp} takes advantage of this symmetry to transform the matrices $\boldsymbol{\Gamma}$,  $\boldsymbol{\mathcal{H}}_\text{int}$, $ \boldsymbol{\cal D}^{nm}_q$, $\boldsymbol{\cal D}^{*,nm}_q$, and $\boldsymbol{\cal M}_{n,q}$ into evolution for the real and imaginary parts of $\rho_{ij}$, reducing the effective number of coupled, first order differential by a factor of two and increasing computational speed.

The particle's semiclassical motion can be calculated through
\begin{equation}
	\ddot{\mathbf{r}} = -\frac{1}{M} \left\{\nabla H\right\} + \mathbf{a} = \frac{1}{M}\left\{\nabla (\mathbf{d}^{nm} \cdot \mathbf{E}_{0,n\rightarrow m} + \boldsymbol{\mu}_n\cdot \mathbf{B})\right\} + \mathbf{a},
\end{equation}
where $\mathbf{r}$ is the position of the atom or molecule, $M$ is its mass, and $\mathbf{a}$ is a constant acceleration (typically gravity).
The motion is also impacted by the momentum diffusion tensor, $D_{ij}=d/dt(\langle P_i P_j \rangle - \langle P_i \rangle \langle P_j \rangle$,
where the $i$ and $j$ indices here represent spatial coordinates and $P$ is the momentum of the atom.
Evaluation of the momentum diffusion tensor is a complicated calculation that involves multiple commutators of the $\mathbf{d}^{nm}$ operators.
The result can be broken into two components: one component that depends on the history of the force applied (the stimulated emission component) and one that depends on the instantaneous populations of the excited states (the spontaneous emission component).
We neglect the former and  generalize the result of Ref.~\cite{Gordon1980} for the latter, so
\begin{equation}
    \label{eq:momentum_diffusion}
    2 D_{ii} = k_{n\rightarrow m}^2 \sum_{n<m} \Gamma^{m\rightarrow n}_{i} \rho^{mm}_{ii},
\end{equation}
where $\Gamma^{m\rightarrow n}_i$ is the decay rate of state $i$ out of manifold $m$ into lower manifold $n$ ($\Gamma^{m\rightarrow n}_i$ is the appropriate diagonal element of the $\boldsymbol{\Gamma}^{m\rightarrow n}$ matrix).
(We neglect the small directional dependence evaluated in Ref.~\cite{Ungar1989}.)
To simulate the effect of the momentum diffusion tensor, {\tt pylcp} uses a Monte-Carlo method.
During integration of the atomic motion, the decay probability in a time step $\Delta t$ is calculated to be $\Delta t \Gamma^{m\rightarrow n}_i \rho^{mm}_{ii}$.
A pseudo-random number generator then generates a random number $p$, which is uniformly distributed on the interval (0, 1].
If $p<\Delta t \Gamma^{m\rightarrow n}_i \rho^{mm}_{ii}$, two randomly directed momentum kicks of $2\hbar k$ are delivered to the atom in order to reproduce the diffusion \eqref{eq:momentum_diffusion}.
The integrator used has an adaptive time step, but it is constrained such that $\sum_{inm}\Gamma^{m\rightarrow n}_i \rho^{mm}_{ii}  \Delta t < 0.1$ by default, minimizing the probability of multiple scattering events per time step.

\subsubsection{Representation of the Hamiltonian}
With the basis states $\ket{i,n}$ arranged in a vector by increasing state and manifold number, the Hamiltonian becomes a $N_m\times N_m$ matrix of submatrices, where $N_m$ is the number of manifolds.
In {\tt pylcp}, we represent this Hamiltonian as a series of blocks
Using labels $i=g,e$ for the two extreme manifolds, the Hamiltonian blocks look like
\begin{equation}
    \label{eq:ham_matrix_form}
    H_\text{atom} = \left(
    \begin{array}{ccc}
    (H_g - \boldsymbol{\mu}_g\cdot \mathbf{B}) & \cdots & (\boldsymbol{d}^{ge}\cdot\mathbf{E}_{ge}^*) \\
    \vdots & \ddots & \vdots  \\
    (\boldsymbol{d}^{*ge}\cdot\mathbf{E}_{ge}) & \cdots & (H_e+\boldsymbol{\mu}_e\cdot \mathbf{B})
    \end{array}\right),
\end{equation}
where each element in the matrix $H_g$, $H_e$, etc. is itself a matrix.
With this structure, the term $\boldsymbol{\mu}_n\cdot\mathbf{B}$ is the field dependent term that mixes states within a given manifold $m$ and $\boldsymbol{d}^{mn}\cdot\mathbf{E}_{mn}$ is the field dependent term that couples states of manifolds $m$ and $n$.

To specify the Hamiltonian, the user provides the requisite $H_n$, $\boldsymbol{\mu}_n$, and $\mathbf{d}^{nm}$ and combines them together in the {\tt hamiltonian} class.
The class stores the block structure of the Hamiltonian, and also contains methods useful for its manipulation.
Because the $\mathbf{d}^{nm}$ components can create photons, the user can specify the associated $k$ and $\Gamma$ with each $\mathbf{d}^{nm}$.
The vectors $\boldsymbol{\mu}$ and $\boldsymbol{d}$ are represented in spherical polar coordinates, allowing for easy translation into $\sigma^{\pm}$ circular polarization and $\pi$ polarizations.
For these vectors, we denote the component $q=\pm1, 0$ as $d_q$.

\subsubsection{Fields}
The electric field for an individual laser, labeled by $l$, is assumed to have the form
\begin{equation}
	\mathbf{E}_l = \frac{1}{2}\hat{\boldsymbol{\epsilon}_l}(r, t) E_l(r, t)e^{i\mathbf{k}_l(r,t)\cdot\mathbf{r}-i \int dt\Delta_l(t) + i\phi_l(r, t)},
\end{equation}
where the complex conjugate term is neglected by the rotating wave approximation.
The user specifies all components of the laser field: the polarization vector $\hat{\boldsymbol{\epsilon}}_l(r, t)$, the amplitude $E_{l}(\mathbf{r}, t)$ (defined through the saturation intensity $s_l(\mathbf{r},t)$), the $\mathbf{k}_l(\mathbf{r},t)$ vector, the average detuning between the rotating frame and optical frequency $\Delta_l(t)$, and any potential phase modulation of the laser beam $\phi_l(t)$ in the {\tt laserBeam} class.
Each of these quantities may be constant or a function of the position $\mathbf{r}$ or time $t$.
Laser beams are collected together into {\tt laserBeams} objects by which manifold transition they address.
If there are multiple manifold transitions, a python dictionary should be made that is indexed by the manifold labels, e.g. $g\rightarrow e$.
Finally, the magnetic field $\mathbf{B}(\mathbf{r}, t)$ can be specified simply as a user-defined function of position $\mathbf{r}$ and/or time $t$.

\subsubsection{Units}
\label{sec:obe_units}
For {\tt pylcp}, $\hbar=1$ so that angular frequencies and energies are equivalent.
Thus, the user specifies $\hat{H}/\hbar$, rather than $\hat{H}$.
For other quantities, we specifically separate out the units, with dimensionless numbers denoted with a bar.
For example, a position $x$ is related to its dimensionless counterpart through $x = \bar{x}x_0$, where $x_0$ is the unit of $x$.
When forming a complete Hamiltonian using the {\tt hamiltonian} class, the user has the ability to set the units by specifying base units of length $x_0$, time $t_0$, magnetic field $B_0$ and mass $m$.

To understand the units associated with $\mathbf{d}\cdot \mathbf{E}$, let us briefly consider a two level system.
The excited state decay rate is given by $\Gamma = k^3(d d^*+d^* d)/6 \pi\epsilon_0\hbar$.
We define a normalized $d=\bar{d} d_0$, where $d_0 = \sqrt{3 \pi \epsilon_0 \hbar \Gamma/k^3}$.
With this definition, $\bar{d}^*\bar{d}+\bar{d}\bar{d}^* = \Gamma/2$.
For the electric field, we use the on-resonant, two-level saturation parameter $s = I/I_\text{sat} = 2|d_0 E|^2/\hbar^2|\gamma|^2 = 8 d_0 |E|^2/\hbar^2\Gamma^2$, where $\gamma = \Gamma/2$ to define the natural units for $E$.
Inverting, we find that $E = \hbar \Gamma \sqrt{s}/(\sqrt{8} d_0)$, so $dE/\hbar = (\bar{d} \Gamma/4) \sqrt{2 s}$.
We note that this definition of the saturation parameter is consistent with the more common $I/I_\text{sat} = 2(\Omega^2/\Gamma^2)$, where, because the Hamiltonian is defined without a factor of 1/2 in the $\mathbf{d}\cdot \mathbf{E}$ term, $\Omega/2=dE$.

For the force, we have one additional unit to specify, the mass.
If the user specifies the length, time, and magnetic field, then
\begin{equation}
    \frac{x_0}{t_0^2} \ddot{\bar{\mathbf{r}}} = \frac{\hbar}{x_0 t_0 M}\left\{ -\bar{\nabla} \bar{H}\right\} + \frac{x_0}{t_0^2}\bar{\mathbf{a}}.\label{eq:motion}
\end{equation}
The default unit selection for a two-manifold system is $t_0=1/\Gamma$ and $x_0=1/k$, which yields for the prefactor $\hbar/x_0 t_0 M = \hbar k \Gamma/M$.
Taking into account the units on the left hand side of~\eqref{eq:motion}, we define
\begin{equation}
	\ddot{\bar{\mathbf{r}}} = \frac{\hbar t_0}{x_0^2 m}\left\{ -\bar{\nabla} \bar{H}\right\} + \bar{\mathbf{a}},
\end{equation}
which defines the `dimensionless' mass $\bar{M} = x_0^2 M/\hbar t_0$.

\subsection{The Rate Equations}
In the rate equation model, we neglect the coherences in the OBEs and instead focus exclusively on the populations in the state $i$ in manifold $n$, $N^n_i = \rho^{nn}_{ii}$.
Here, we follow the construction of the rate equations from Ref.~\cite{Tarbutt2015}.
The evolution of $N^n_i$ is given by
\begin{equation}
    \dot{N}^n_i = \pm\sum_{m,j,l} R^{n\rightarrow m}_{ij,l} (N^n_i - N^m_j) + \sum_{m>n
    } \Gamma^{m\rightarrow n}_{ij} N^m_j - \sum_{m<n}\Gamma^{n\rightarrow m} N^n_i,
\end{equation}
where the first term accounts for optical pumping, the second for decays into the state, and the third decays from the state.
The $+$($-$) sign occurs when $m<n$ ($n>m$).
The manifold decay rate is
\begin{equation}
	\Gamma^{m\rightarrow n} = \frac{k^3_{n\rightarrow m}}{3\pi\epsilon_0\hbar}|\mathbf{d}^{nm}|^2
\end{equation}
and the decay rate out of the excited state $\Gamma$, and the branching ratio,
\begin{equation}
    \Gamma_{ij}^{n\rightarrow m} = \Gamma \frac{\left|\mathbf{d}^{nm}_{ij}\right|^2}{\sum_i \left|\mathbf{d}^{nm}_{ij}\right|^2}.
\end{equation}
We calculate the optical pumping rates $R_{ij, l}^{n\rightarrow m}$, due to the laser $l$ between states $i$ and $j$ in manifolds $n$ and $m$, respectively.
It is given by
\begin{equation}
    \label{eq:rate_eq:pumping_rate}
    R_{ij,l}^{n\rightarrow m} = \frac{[\Omega^{n\rightarrow m}_{ij,l}]^2/\Gamma^{n\rightarrow m}}{1 + 4[(\Delta_l - (\omega^m_j-\omega^n_i) - \mathbf{k}_l\cdot \mathbf{v})/\Gamma^{m\rightarrow n})]^2},
\end{equation}
where the excitation rate is\footnote{Note that there is a factor of two difference between $d\cdot E$ in Sec.~\ref{sec:obe_units} and the definition of the Rabi rate $\Omega$ used here.}
\begin{equation}
	\label{eq:rabi_rate_rate_eq}
	\Omega_{ij,l}^{n\rightarrow m} = \frac{\Gamma^{m\rightarrow n}}{2} (\mathbf{d}_{ij}^{nm}\cdot \boldsymbol{\epsilon}'_l) \sqrt{2s(\mathbf{r},t)},
\end{equation}
$\omega^m_j$ and $\omega^n_i$ are the eigenenergies of states $\ket{j,n}$ and $\ket{i,n}$, and $\boldsymbol{\epsilon}'_l$ is polarization of laser $l$, rotated into the local coordinate system defined by the quantization axis.
In this approximation, the quantization axis is assumed to be defined by the magnetic field.
To determine $\omega^m_j$ and $\omega^n_i$, each diagonal block of the Hamiltonian \eqref{eq:ham_matrix_form} is diagonalized at each new magnetic field $B$ then the $\mathbf{d}^{nm}$ matrices are rotated accordingly into the new basis.
{\tt pylcp} automatically diagonalizes and rotates these matrices as the atom moves through the magnetic and laser fields.

To determine $\boldsymbol{\epsilon}'_l$, the polarization of the laser fields $\boldsymbol{\epsilon}_l$ must be rotated onto a possibly changing quantization axis.
By default, {\tt pylcp} stores the polarization of a given laser beam as a vector in the spherical basis with $\hat{z}$ assumed to be the quantization axis.
To rotate to a new quantization axis defined by the local magnetic field $\hat{z}'$, we need to apply a Wigner rotation matrix ${\mathscr D}^j_{m'm}$ for an $j=1$ angular momentum system (e.g., a photon).
The specific form is given by
\begin{equation}
  \mathscr{D} = \left(\begin{array}{ccc}
  \frac{1+\cos\beta}{2}e^{-i\alpha+i\gamma} & \frac{\sin\beta}{\sqrt{2}}e^{-i\alpha} & \frac{1-\cos\beta}{2}e^{-i\alpha-i\gamma} \\
  -\frac{\sin\beta}{\sqrt{2}}e^{i\gamma} & \cos\beta & -\frac{\sin\beta}{\sqrt{2}}e^{-i\gamma} \\
  \frac{1-\cos\beta}{2}e^{i\alpha+i\gamma} & \frac{\sin\beta}{\sqrt{2}} & \frac{1+\cos\beta}{2}e^{i\alpha-i\gamma}
  \end{array}\right),
\end{equation}
where $\alpha$, $\beta$, and $\gamma$ in this context are the Euler angles in the $ZYZ$ convention.
Thus, $\boldsymbol{\epsilon}'_l = \mathscr{D}\cdot \boldsymbol{\epsilon}_l$.
For the rate equations, which square the rotated polarization, the angle $\alpha$ does not contribute, so {\tt pylcp} neglects it ($\alpha$ only specifies the relative phase between the two circular components in the rotated coordinate system).

Finally, the motion of the atom is given by
\begin{equation}
	\mathbf{\ddot{r}} = \sum_{l} \frac{\hbar \mathbf{k}_l}{2 M}\sum_{m, n, i,j} R_{ij,l}^{n\rightarrow m}(N^m_j - N^n_i) + \mathbf{a}.
\end{equation}
Simulated spontaneous emission follows in the same way as in the OBEs.

\subsection{Heuristic equation}
A final governing equation is also included in {\tt pylcp}, titled the `heuristic' equation.
This equation calculates the force on an atom assuming a $F=0\rightarrow F'=1$ level structure.
This level structure has equal dipole transition strengths for all three transitions, making the $(\mathbf{d}_{ij}^{nm}\cdot \boldsymbol{\epsilon}_l) = \epsilon_l$ and
\begin{equation}\label{eq:heuristic}
	\ddot{\mathbf{r}} = \frac{\hbar \mathbf{k} \Gamma}{2 M}\sum_{l,q} \frac{s_l \left({\epsilon}'_{l,q}\right)^2}{1+\sum_j s_j+4(\Delta^2 - \mathbf{k}\cdot\mathbf{v} - q |B|)^2/\Gamma^2},
\end{equation}
where ${\epsilon}'_{l,q}=\left(\mathcal{D}\cdot\boldsymbol{\epsilon}_l\right)_q$ is the $q$th component of the rotated polarization $\boldsymbol{\epsilon}'_l$.
Here, we have approximated the total saturation as $\sum_j s_j(\mathbf{r})$, which is the approximation used in Ref.~\cite{Lett1989}.

\section{Examples}
\label{sec:examples}
In this section, we cover several  examples of laser cooling and show that we can use {\tt pylcp} to reproduce standard results.
The examples described herein are a small subset of the examples contained within the {\tt pylcp} package.
All examples are included in the user manual and are saved as {\tt jupyter} notebooks that include comments and saved output figures.

\subsection{Stationary atoms: internal dynamics}\label{sec:internal}
\begin{figure}
	\center
	\includegraphics{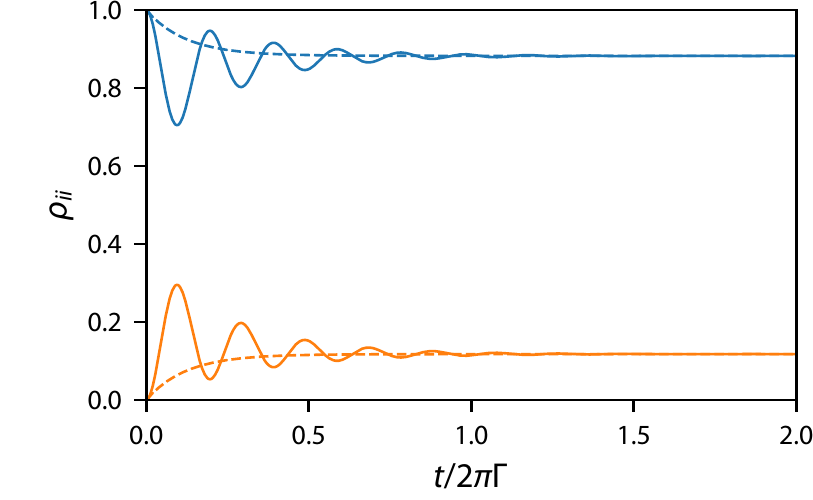}
	\caption{\label{fig:rabi_flopping}
	Populations of a two level atom $\rho_{00}$ (blue) and $\rho_{11}$ (orange) vs. time $t$ showing damped Rabi flopping with both the OBEs (solid) and the rate equations (dashed).
	Here, the detuning is $\Delta/\Gamma=-4$ and intensity is $s_0=20$.}
\end{figure}

We start by considering well-known examples of internal atom dynamics~\cite{Foot2005, Metcalf1999}.
These examples are contained in the {\tt basics} subdirectory of the {\tt pylcp} package.
Our first example is damped Rabi flopping.
When a two-level atom (states $\ket{0}$ and $\ket{1}$) is illuminated with off-resonant laser light with detuning $\Delta=-4\Gamma$ and saturation $s_0 = 20$, it drives oscillations in the population of the atomic states.
The decay of the excited state leads to the oscillation decaying and the populations achieving an equilibrium value.
In Fig.~\ref{fig:rabi_flopping}, the populations $\rho_{00}$ and $\rho_{11}$, as determined by both the OBEs and the rate equations, are shown.
Unlike the OBEs, the rate equations show no oscillation.
This is expected as the rate equations neglect the coherences between the atomic states.
Nevertheless, the solutions of both the OBEs and rate equations asymptotically approach the same equilibrium populations.

\begin{figure}
	\center
	\includegraphics{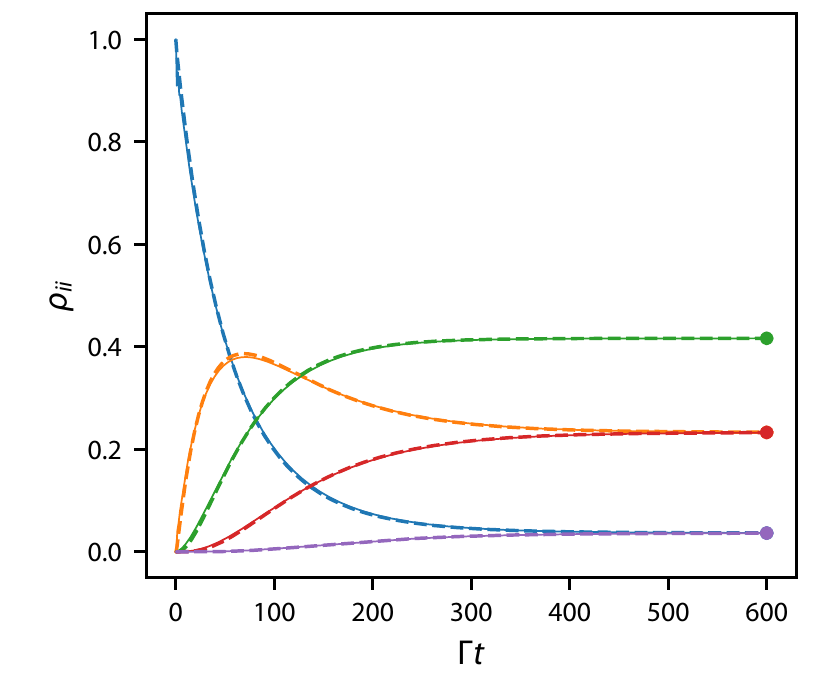}
	\caption{\label{fig:optical_pumping}
	Optical pumping on an $F=2\rightarrow F'=3$ transition of an atom.  The populations of the $m_F=-2, 1, 0, 1, 2$ states are shown in blue, orange, green, red, and purple, respectively.  The solution from the rate equations (OBEs) are shown as dashed (solid) curves.  See text for other parameters.}
\end{figure}

Next, consider a more complicated $F=2\rightarrow F'=3$ atom initialized in $\ket{F=2,m_F=-2}$ and illuminated by $\pi$-polarized light with respect to the quantization axis, \(\hat{z}\).
In this configuration, the laser optically pumps the atoms into a stretched state with $\langle F_z \rangle = 0$, corresponding to equal populations between $\pm m_F$ with maximal population in $m_F=0$.
The optical pumping process is depicted in Fig.~\ref{fig:optical_pumping}, using a detuning $\Delta/\Gamma = -2.73$ and $s_0= 1.352464$, which match the parameters used for an equivalent simulation in Ungar, {\it et. al.}~\cite{Ungar1989}\footnote{Note that $s_0$ therein is defined with the detuning included.}.
Fig.~\ref{fig:optical_pumping} shows the solutions of both the OBEs and the rate equations, which are nearly identical.

Other two-level examples contained in {\tt basics} subdirectory of the {\tt pylcp} package, but omitted from the discussion here include power broadening, undamped Rabi flopping in a magnetic field, and optical pumping of an $F=1\rightarrow F'=2$ transition in the presence of a magnetic field.
These examples and the ones discussed above all use fields that are constant in time.
To demonstrate its applicability to time varying fields, {\tt pylcp} also includes an example of adiabatic rapid passage that reproduces Fig.~2 of Ref.~\cite{Lu2005}.

\begin{figure}
	\center
	\includegraphics{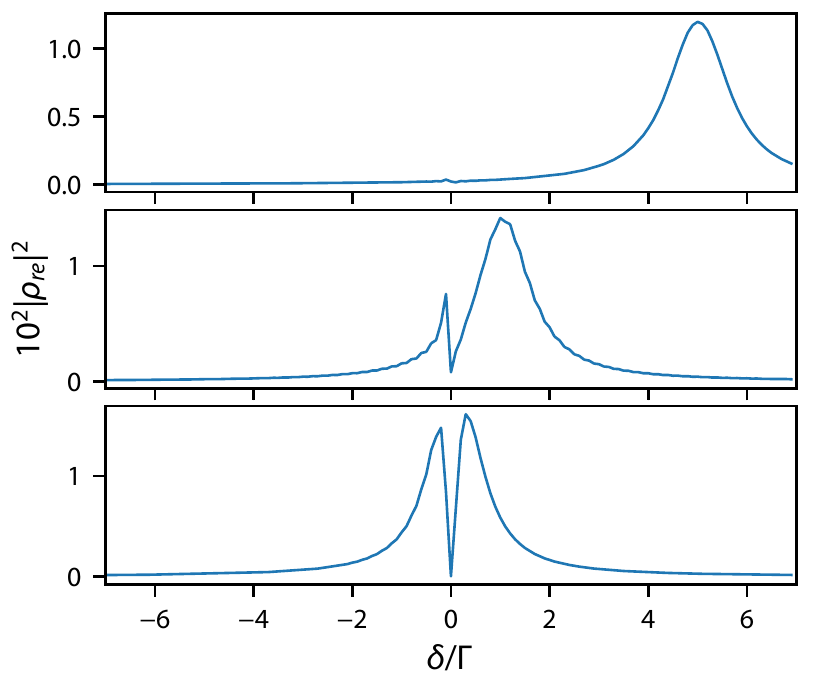}
	\caption{\label{fig:eit_susceptibility} Susceptibility ($\rho_{re}$) of a three-level atom as a function of $\delta = \Delta_{ge}-\Delta_{re}$ for $\Delta_{ge}/\Gamma=-5$ (left), $-1$ (middle), and $-0.1$ (right).}
\end{figure}

Beyond the two-level systems considered above, a three-level atom can produce a multitude of more complicated physical phenomena.
One well-known example is electromagnetically induced transparency.
Consider a three-level atom in the $\Lambda$ configuration, with ground state $\ket{g}$, intermediate metastable state $\ket{r}$, and excited state $\ket{e}$.
Two lasers, with Rabi frequencies \(\Omega_{ge}\) and \(\Omega_{re}\), drive the \(\ket{g}\rightarrow\ket{e}\) and \(\ket{r}\rightarrow\ket{e}\) transitions, respectively.
We assume $\Omega_{ge} = \Braket{g|\mathbf{d}^{ge}\cdot \mathbf{E}_{0,g\rightarrow e}|e} \gg \Omega_{re} = \Braket{r|\mathbf{d}^{re}\cdot \mathbf{E}_{0,r\rightarrow e}|e}$.
In this limit, when both lasers are resonant with their respective transitions, the strong \(\ket{g}\rightarrow\ket{e}\) laser opens a transparency window in the otherwise absorptive \(\ket{r}\rightarrow\ket{e}\) transition~\cite{Boller1991}.
This transparency window can be seen by observed in the susceptibility through the coherence $\rho_{re}$.
Figure~\ref{fig:eit_susceptibility} shows the numerically calculated $\rho_{re}$ as a function of $\delta/\Gamma = \Delta_{ge}/\Gamma-\Delta_{re}/\Gamma$ for three values of $\Delta_{ge}/\Gamma$, where $\Delta_{ie}$ is the detuning of the laser driving the $\ket{i}\rightarrow\ket{e}$ transition from resonance.
The Rabi frequencies for the two transitions are given by $2(\Omega_{ge}/\Gamma)^2=10$ and $2(\Omega_{re}/\Gamma)^2=0.1$.
For $|\Delta_{ge}|\gg 1$, the $\ket{r}\rightarrow\ket{e}$ absorption profile is not perturbed by the presence of the coupling laser driving $\ket{g}\rightarrow\ket{e}$, except for a small glitch in the susceptibility near Raman resonance at $\delta/\Gamma=0$.
As $|\Delta_{ge}|\rightarrow 0$, the susceptibility drops to zero when $\delta/\Gamma=0$ and the atom no longer absorbs photons from the $\ket{r}\rightarrow\ket{e}$ laser.

\begin{figure}
	\center
	\includegraphics{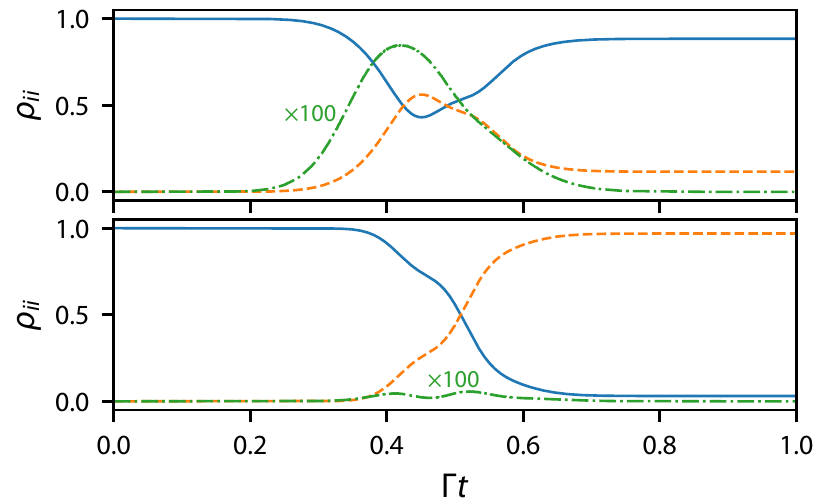}
	\caption{\label{fig:stirap} STIRAP time dynamics of a $\Lambda$ system with state populations $\rho_{gg}$ (solid blue), $\rho_{rr}$ (dashed orange) and $\rho_{ee}$ (dashed-dot green) with Gaussian-shaped pulses (see text).
	Top (bottom) shows the incorrect (correct) pulse order.}
\end{figure}

Another interesting three-level effect is stimulated Raman adiabatic passage (STIRAP)~\cite{Bergmann1998}.
STIRAP is an adiabatic process that transfers population between two states in a three-level system.
Consider the $\Lambda$ system defined above in the electromagnetically induced transparency discussion.
Counterintuitively, efficient population transfer from $\ket{g}$ to $\ket{r}$ requires the optical fields to first couple only $\ket{r}$ to $\ket{e}$ and then slowly change optical intensities until only $\ket{g}$ is coupled to $\ket{e}$.
Fig.~\ref{fig:stirap} shows a {\tt pylcp} calculation of STIRAP, using pulses with Gaussian temporal profiles: $\Omega_{ge}(t)=\Omega_{ge,0}e^{-2(t-t_0+\Delta t/2)^2/t_w^2}$ and $\Omega_{re}(t)=\Omega_{re,0}e^{-2(t-t_0-\Delta t/2)^2/t_w^2}$ with $\Omega_{ge,0}=\Omega_{re,0}=\Gamma$ and $t_0=500/\Gamma$.
For the correct pulse order $\Delta t=125/\Gamma$ (bottom panel), we see efficient population transfer from $\ket{g}$ to $\ket{r}$ with minimal population $\rho_{ee}$ in $\ket{e}$.
For the incorrect pulse order $\Delta t=-125/\Gamma$ (top panel), we see the opposite.

\subsection{One-dimensional optical molasses}
The examples in Sec.~\ref{sec:internal} exclusively examined the internal structure and dynamics of an atom.
However, the {\tt pylcp} package is focused on simulating the coupling of internal atomic dynamics to external atomic motion.
One well-understood problem in this realm is one-dimensional optical molasses, where a two-level atom moves through an optical field generated by two counter-propagating laser beams~\cite{Lett1989}.

\begin{figure}
	\center
	\includegraphics{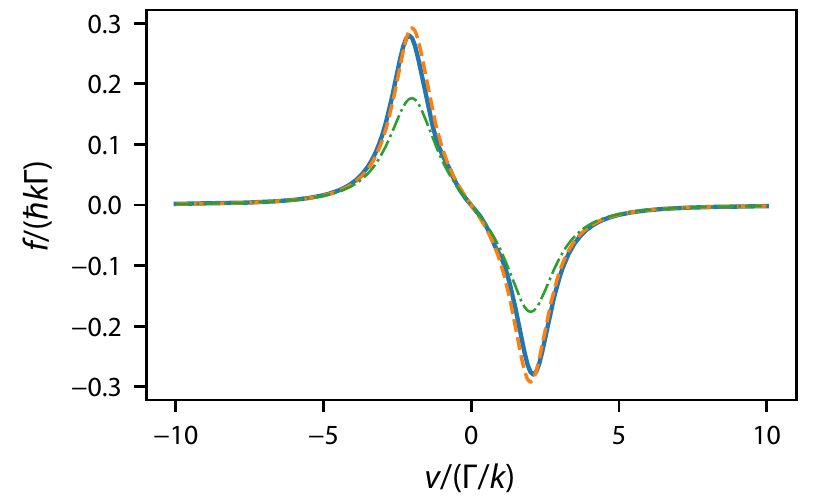}
	\caption{\label{fig:two_level_molasses_forces} Force $f$ as a function of velocity $v$ in a one-dimensional two-level molasses.
	The three curves show the differences between the OBE (blue, solid), rate equations (orange, dashed), and heuristic equation (green, dot-dashed).}
\end{figure}

We begin by calculating the equilibrium force on the atom in the optical molasses as a function of velocity~\cite{Chu1985, Lett1989}.
Fig.~\ref{fig:two_level_molasses_forces} shows the force calculated by {\tt pylcp} using our three governing equations (see Sec.~\ref{sec:governing_equations}) with $\Delta/\Gamma=-2$ and $s=1.5$.
Equilibrium forces in an optical molasses are most accurately described using either the OBEs or the rate equations, which accurately account for saturation from the beams for all velocities.
The heuristic equation overestimates the saturation except near the origin, causing an underestimation of the force when $|v|\gg \Gamma/k$.
The damping force near the origin is nearly identical between the three governing equations when $\Delta/\Gamma$ is small.

\begin{figure}
	\center
	\includegraphics{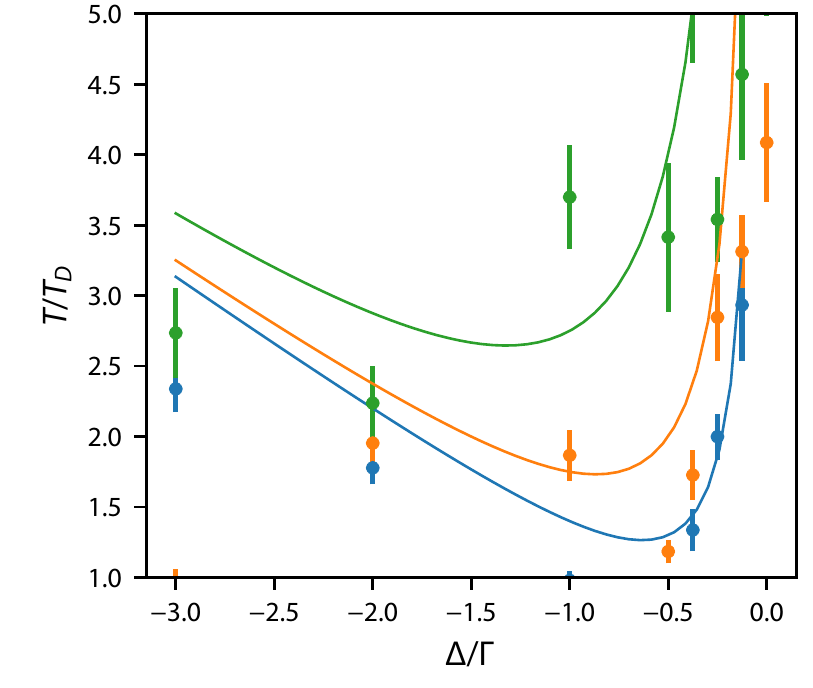}
	\caption{\label{fig:doppler_limit}
	Final temperature $T$ vs. detuning $\Delta$ and $s_0=0.3,\ 1,\ 3$ (blue, orange, green, respectively) for two-level atoms in a 1D molasses.
	Points show rate equation simulations, with error bars denoting 1-$\sigma$ uncertainty of 100 simulations.
	Lines denote analytic Doppler temperature of~\eqref{eq:general_doppler_limit}.}
\end{figure}

Figure~\ref{fig:doppler_limit} shows the temperatures of small clouds ($N=100$ atoms) simulated with the rate equations in a 1D molasses for $t_\text{max} = 10^3/\Gamma$.
We see reasonably good agreement with the expectation of Doppler cooling theory, Eq.~(18) in Ref.~\cite{Lett1989}.
Specifically, the solid curves show the expected value of
\begin{equation}
	\label{eq:general_doppler_limit}
	\frac{T}{T_D} = \frac{1+2s_0+4(\Delta/\Gamma)^2}{4|\Delta/\Gamma|},
\end{equation}
where $T_D = \hbar \Gamma/2k_B$.
We note that we generally observe a lower temperature than that predicted by \eqref{eq:general_doppler_limit} because the heuristic equation tends to underestimate the damping at larger detuning, and \eqref{eq:general_doppler_limit} comes from an analysis of the heuristic equation.

\begin{figure}
	\center
	\includegraphics{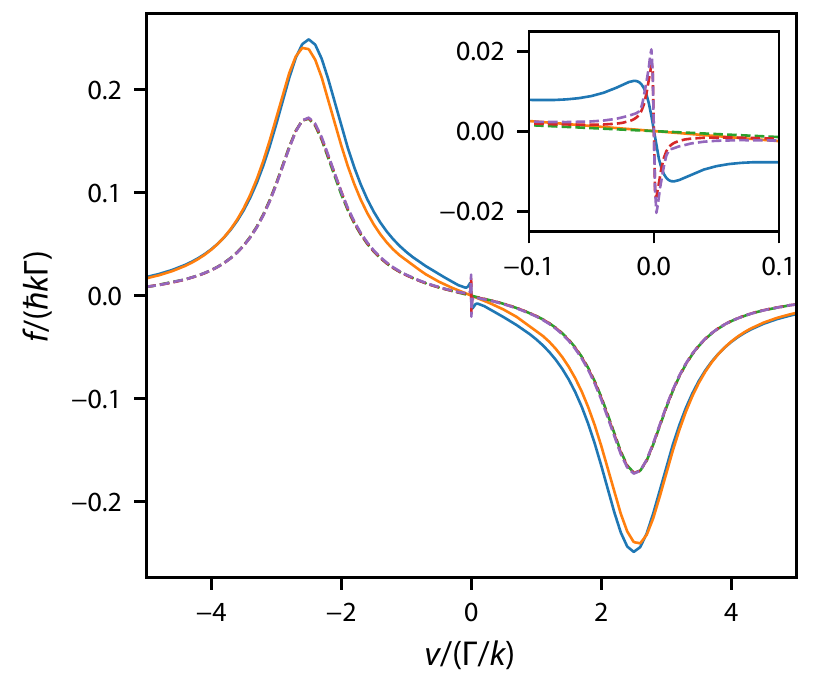}
	\caption{\label{fig:sub_doppler_force} Force $f$ vs. velocity $v$ in a one-dimensional $F=2\rightarrow F'=3$ molasses with $\Delta/\Gamma=-2.5$ and $s_0=1.0$.  The solid curve correspond to circularly polarized beams: $\sigma^+\sigma^+$ (orange) and $\sigma^+\sigma^-$ (blue).  Dashed lines show linearly-polarized input beams with angles $\phi=0$ (green), $\phi=\pi/4$ (red), $\phi=\pi/2$ (purple). Inset shows features close to the origin.}
\end{figure}

We can extend this study of one-dimensional molasses with more complicated level structures.
In particular, we focus on the $F=2\rightarrow F'=3$ transition in $^{23}$Na, which was studied first by Ungar, {\it et. al.}~\cite{Ungar1989}.
Figure~\ref{fig:sub_doppler_force} shows the calculated equilibrium force vs. velocity and contains all of the essential features that were discussed in Ref.~\cite{Ungar1989}.
First, linear polarization $\phi\neq 0$ and the so-called `corkscrew' polarization $\sigma^+\sigma^-$ produce sub-Doppler force features near zero velocity that create additional damping.
Second, compared to the `two state' $\sigma^+\sigma^+$ polarization, $\sigma^+\sigma^-$ produces excess force due to the sub-Doppler discontinuity at $v=0$ that continues out to near $v/k\approx \Delta$, where one of the two lasers is dominantly resonant with the atom.
In the example, but omitted here, we also replicate the dependence of the force on saturation $s_0$ and detuning $\Delta$ observed in Ref.~\cite{Ungar1989}.

In addition to the static force profiles, we also replicate the temperature observed via Monte-Carlo in Ref.~\cite{Ungar1989}.
In particular, we have simulated 96 atoms in the $\sigma^+\sigma^-$ configuration for a time $t=10^4/\Gamma$.
We sample the atoms' velocity at a time interval $\delta t=500/\Gamma$ after allowing them to come into equilibrium with the light field for $2000/\Gamma$.
Fitting the resulting sampled velocities to a Gaussian reveals a temperature of $6.1(1.5)$~$\mu$K.
This temperature compares nicely to the 8.2~$\mu$K obtained in Ref.~\cite{Ungar1989}.
We note that our temperature should be lower than that of Ref.~\cite{Ungar1989} as we do not include contributions to the momentum diffusion tensor from stimulated emission, for which they include an approximate contribution.

Contained in the examples directory, but omitted here, is a calculation of one-dimensional molasses for a variety of different polarizations and level structures, including those that utilize type-II transitions that have dark states.
Molasses operating on these transitions was studied in Ref.~\cite{Devlin2016} and the essential results are reproduced in the examples.

Another type of sub-Doppler molasses that is useful is that of three-level $\Lambda$-enhanced cooling~\cite{Grier2013}.
This molasses technique is unique compared to the others described above in that it uses a three-manifold system rather than a two-manifold.
We have simulated this process using {\tt pylcp}, and the script and results are contained with the other molasses examples.
The numerical results match the theoretical results of Ref.~\cite{Grier2013}.

\subsection{Force, capture, and temperature in MOTs}

\begin{figure}[t]
	\center
	\includegraphics{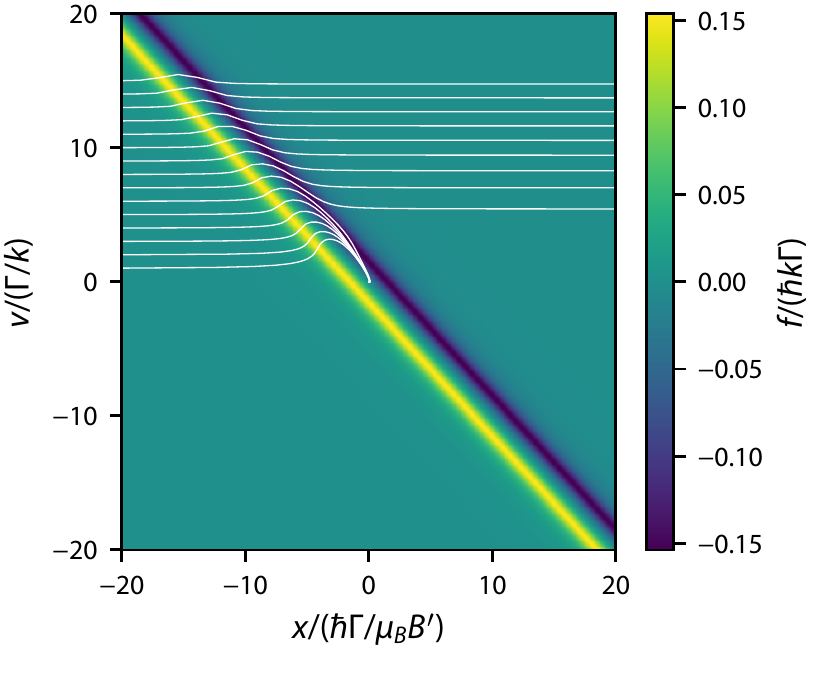}
	\caption{\label{fig:mot_forces} Force $f$ vs. both position $x$ and velocity $v$ in a one-dimensional magneto-optical trap with laser parameters $\Delta/\Gamma=1.5$ and $s_0=1$.  The white curves show trajectories through phase space for atoms entering the MOT with different initial velocities $v_0$.}
\end{figure}

\begin{figure}
    \centering
    \includegraphics{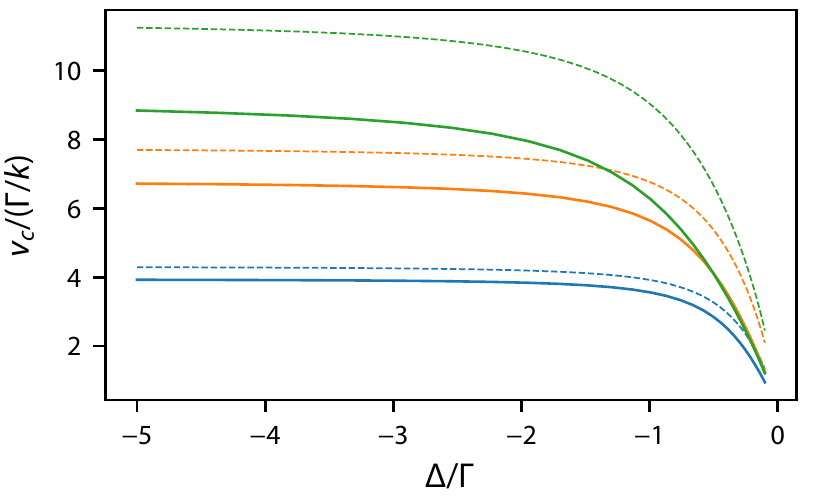}
	\caption{\label{fig:mot_capture} Capture velocity $v_c$ vs. laser detuning $\Delta$ for $s_0=0.3$ (blue), $1.0$ (orange), and $3.0$ (green). Solid curves are calculated using the heuristic equation \eqref{eq:heuristic}; dashed curves are the theoretical result from Ref.~\cite{Haubrich1993}.}
\end{figure}

Forces in a magneto-optical trap (MOT) depend on position as well as velocity.
Let us consider the one-dimensional MOT, where two lasers, one travelling along $+\hat{x}$ with circular polarization $\sigma^-$ (relative to the quantization axis +$\hat{x}$) and the other travelling along $-\hat{x}$ with polarization $\sigma^+$.
These lasers interact with an $F=0\rightarrow F'=1$ atom in a magnetic field of the form $\mathbf{B} = B'\hat{x}$.
Figure~\ref{fig:mot_forces} shows the resulting forces, calculated using the heuristic equation \eqref{eq:heuristic} for a one-dimensional MOT.
Here, we use laser parameters $\Delta/\Gamma=-1.5$ and $s_0=1$ and measure lengths in terms of the Zeeman detuning, $\hbar\Gamma/(\mu_B B')$.
A particle initially at the origin of this non-conservative trap will be forced back toward the origin if it gains any velocity or is displaced.

The capture process into such a MOT can easily be simulated with {\tt pylcp}.
The white curves in Fig.~\ref{fig:mot_forces} show trajectories in the classical phase space, calculated without random scattering.
Here, atoms enter the MOT at a position $x=-20(\hbar \Gamma/\mu_B B')$ with varying initial velocity $v_0$.
Clearly, trajectories with $v_0 < 7\Gamma/k$ are damped to the origin and captured, whereas trajectories with $v_0\geq 7 \Gamma/k$ escape.
Employing a binary search algorithm, one can obtain the MOT capture velocity numerically.
Figure~\ref{fig:mot_capture} shows the result of such a calculation using the heuristic equation \eqref{eq:heuristic} along with the prediction of a simple model found in Ref.~\cite{Haubrich1993}.
The two calculations agree rather well for small $s_0\leq 1$, but diverge by more than 10~\% for larger $s_0>1$.
This procedure has been extended to three-dimensional MOTs with non-standard geometry and should be the subject a forthcoming paper.

The {\tt pylcp} documentation also includes an example OBE calculation of the temperature of a three-dimensional, $F=0\rightarrow F'=1$ magneto-optical trap.
Using $\Delta/\Gamma=-2.5$ and $s_0=1.25$, we simulated the trajectories of 96 atoms for a time $t=10^5/\Gamma$.
By sampling the velocity distribution every $10^4/\Gamma$ for the latter half of the solution, we find a $T/T_D=3.9(1)$.  The example uses the default units described in Sec.~\ref{sec:obe_units} and takes the mass of the atom simulated to be $\bar{M}=100$, closest to $^7$Li which has $\bar{M}\approx 46$.

\begin{figure}[t]
	\center
	\includegraphics{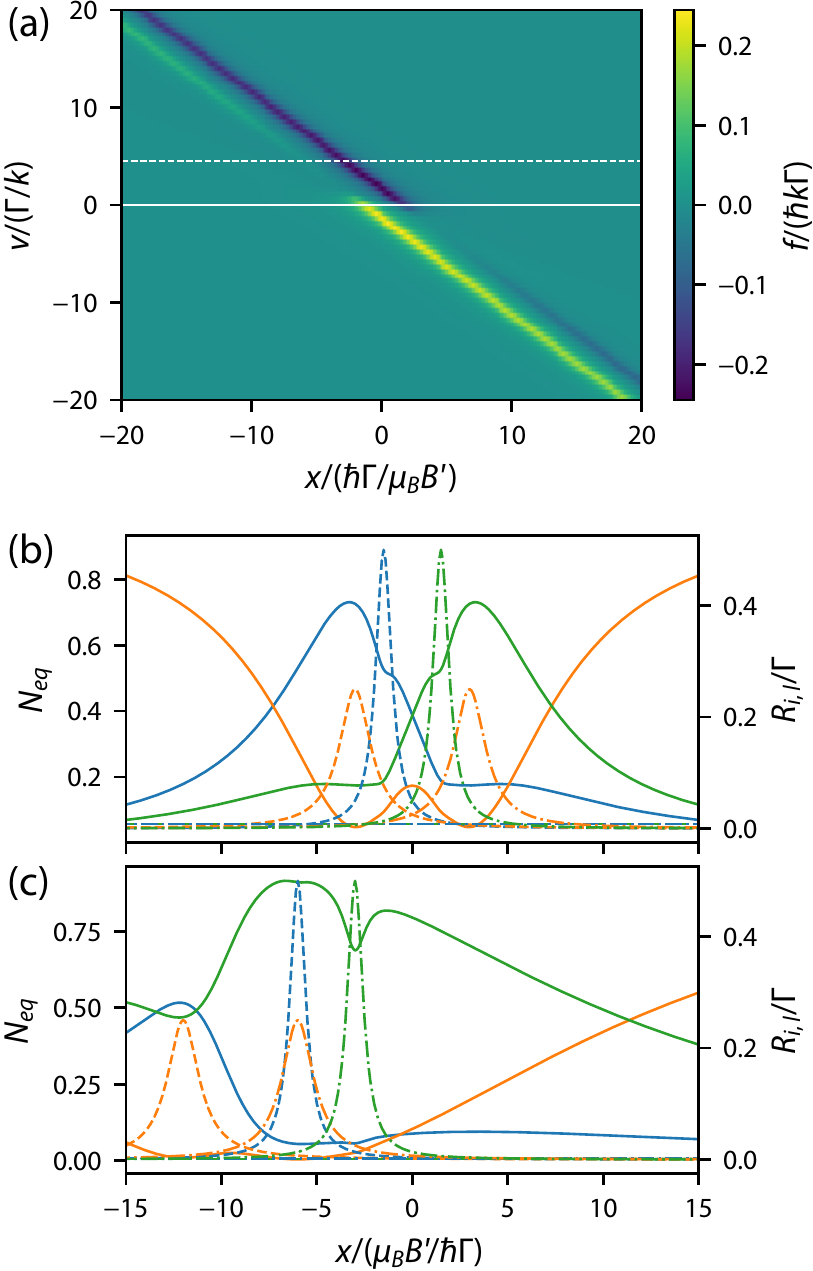}
	\caption{\label{fig:mot_F1_to_F2}
	Anomalous forces in a $F=1\rightarrow F'=2$ MOT with $g_F=0$ and $g_F'=1/F'$.
	(a) Force $f$ vs. position $x$ and velocity $v$ for $\Delta/\Gamma=-4$ and $s_0=5$.
	(b) Equilibrium populations $N_{eq}$ (solid) and pumping rates $R_{i,l}$ from ground state $m_F=-1, 0, 1$ (blue, orange, red, respectively) by the $+\hat{x}$ directed laser (dashed) and the $-\hat{x}$ directed laser (dash-dotted) along the horizontal solid line in (a).
	(c) Same as (b), but along the horizontal dashed line in (a).}
\end{figure}

Extending beyond the $F=0\rightarrow F'=1$ atom  to larger $F$ and $F'$ in the one-dimensional case produces some intriguing results.
Consider Fig.~\ref{fig:mot_F1_to_F2}(a) which shows the force on a $F=1\rightarrow F'=2$ atom, with ground state Land\'{e} \(g\)-factor $g_F=0$ and excited state \(g\)-factor $g_F'=1/F'$, in a one-dimensional MOT.
While not directly applicable to a specific atom, this level structure is a good model of an alkaline earth atom with nuclear spin, where $g_F\approx 0$.
Curiously, the restoring force from the $+\hat{x}$-going beam nearly completely disappears for $v>0$ and vice versa for the beam travelling in $-\hat{x}$.

The loss of restoring force is due to non-cycling transitions becoming resonant through Doppler shifts.
Figure~\ref{fig:mot_F1_to_F2}(b) and Fig.~\ref{fig:mot_F1_to_F2}(c) show the equilibrium populations and pumping rates as a function of position for two different velocities [denoted as the white horizontal lines in Fig.~\ref{fig:mot_F1_to_F2}(a)].
The pumping rates shown are the total pumping rates out of the $m_F=-1,0,+1$ states for both lasers, $R_{i,l} = \sum_j R_{ij,l}$.
For $v=0$, shown in Fig.~\ref{fig:mot_F1_to_F2}(b), the populations and pumping rates are symmetric about $x=0$, and the resulting spin expectation value $\langle F_z \rangle$ is anti-symmetric.
The anti-symmetry of $\langle F_z \rangle$ results from the symmetric nature of the optical pumping.
The pumping rates out of $m_F=\pm 1$ peak at positions given by $x = \pm  \hbar\Delta/\mu_B B'$.
The transitions out of $m_F=\pm 1$ are driven by oppositely directed lasers, resulting in a restoring force.
There is a non-negligible pumping rate out of $m_F=0$ driven by both lasers at larger $|x|$, but the pumping is driven by the correct lasers to produce a restoring force.
For $v=5 \Gamma/k$, however, the optical pumping symmetry is lost, as shown in Fig.~\ref{fig:mot_F1_to_F2}(c).
The pumping rates out of $m_F=\pm1$ peak at $x = \pm  \hbar\Delta/\mu_B B' - kv$, shifting the peaks away from the origin.
By itself, the shift is not sufficient to eliminate the restoring force; however, the transitions out of $m_F=0$ are also shifted.
The resonant location of the transition out of $m_F=0$ by the $-\hat{x}$-going laser overlaps with that of the transition out of $m_F=-1$ from the $+\hat{x}$-going laser.
Rather than cycling between $m_F=-1\rightarrow m_F'=-2$, the atoms are pumped into $m_F=1$, where they stop scattering photons and consequently experience no force.
Repeating the force calculation for progressively larger values of $F$, the effect of non-cycling transitions becomes more drastic, with the force cutting off increasingly sharply around $v=0$.

The force profiles in Fig.~\ref{fig:mot_F1_to_F2}(a) still permit stable MOT operation in equilibrium.
However, in MOTs that operate near the recoil limit, individual atoms may not be optically pumped appropriately as they traverse the MOT, leading to rapid atom loss.
The problem of non-cycling transitions being Doppler shifted into resonance and destroying cycling is a well known in narrow-line MOTs for alkaline earth atoms~\cite{Mukaiyama2003}.
For this reason, narrow-line MOTs of $^{87}$Sr, with $F=9/2$, require a mixing laser to randomize population amongst the ground states of the atom in order to maintain optical cycling~\cite{Mukaiyama2003}.

\begin{figure*}
	\center
	\includegraphics{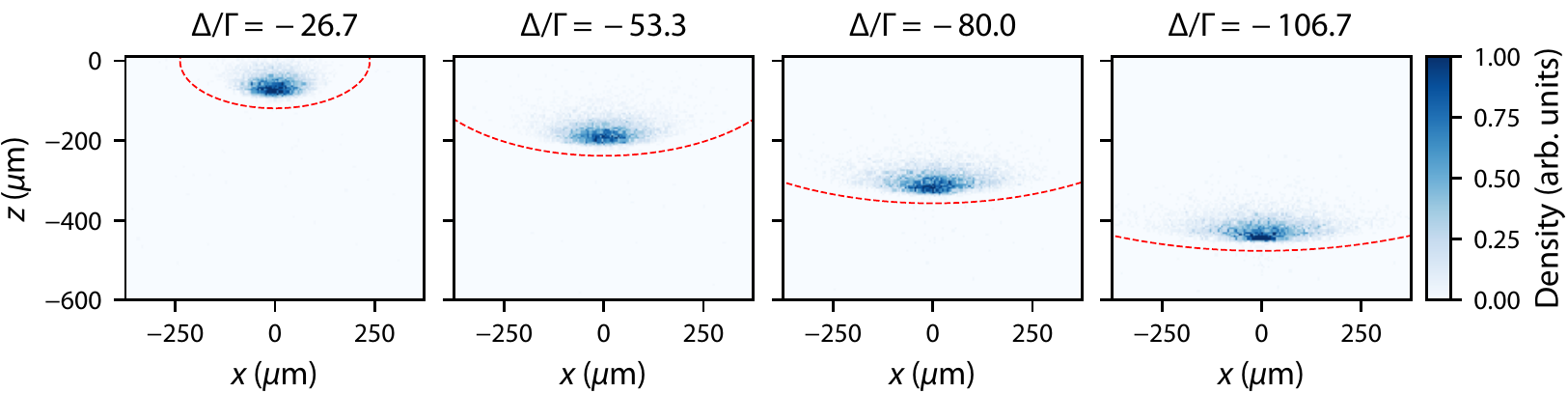}
	\caption{\label{fig:recoil_limited_MOT} Simulated MOT images of the narrow line $^{88}$Sr MOT formed on the $^1 \mbox{S}_{0}\rightarrow\, ^3\mbox{P}_1$ transition for various detunings $\Delta$ and $s_0=25$.  Gravity is in the $-\hat{z}$ direction.  The red, dashed ellipses denote the spatial location where the beam detuning matches the Zeeman shift.  This figure should be compared to Ref.~\cite{Hanley2018}.}
\end{figure*}

Finally, we have reproduced the calculation of Ref.~\cite{Hanley2018} of a recoil-limited MOT.
Using the rate equations with random scattering and the parameters of $^{88}$Sr, we simulate the narrow-line MOT formed on the $^1 \mbox{S}_{0}\rightarrow\, ^3\mbox{P}_1$ transition for $t=50$~ms ($\approx 2400/\Gamma$).
The resulting simulated images of the MOT with $N=1024$ atoms are shown in Fig.~\ref{fig:recoil_limited_MOT}.
One of the most interesting features of narrow-line MOTs is that they sag under the effect of gravity.
The vertical position and shape are determined by the position at which the Zeeman shift compensates for the laser detuning, denoted by the red, dashed ellipses in Fig.~\ref{fig:recoil_limited_MOT}.
One effect included in our simulations that was not discussed in Ref.~\cite{Hanley2018} is atom loss.
With such a large recoil velocity, a few properly directed photon recoils is enough to cause atoms to pass through the region of maximal force and be lost.
We find that, for almost all detunings explored, we lose roughly 3~\% of the atoms during the simulation, corresponding to a loss rate of $0.8(2)$~s$^{-1}$, consistent with experimental results~\cite{Katori1999, Mukaiyama2003,Loftus2004b, Stellmer2013}.

Other MOT examples in the documentation include calculations of MOT damping forces and trapping frequencies (along with comparison to analytic formulas), the forces for MOTs operating on both $D_1$ and $D_2$ lines of alkali atoms~\cite{Flemming1997}, the forces for the variety of different type-I and type-II MOTs considered in Ref.~\cite{Tarbutt2015}, and the forces in the two-color CaF MOT of Ref.~\cite{Tarbutt2015a}.

\subsection{Bi-chromatic forces}

\begin{figure}
	\center
	\includegraphics{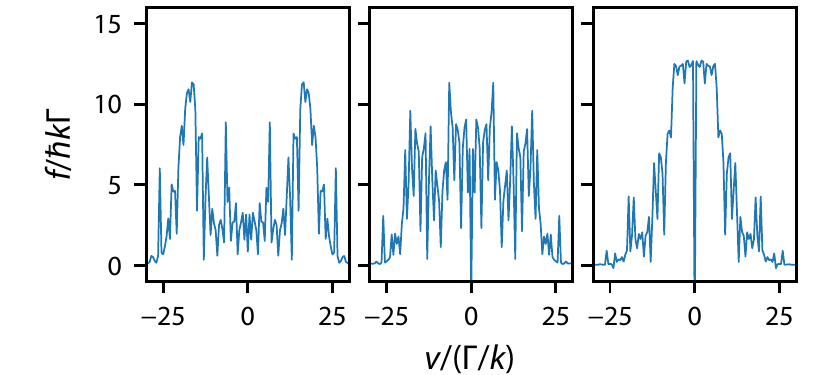}
	\caption{\label{fig:bichromatic} Force $f$ vs. velocity $v$ for a two-level atom in an intense bi-chromatic standing wave for $\Delta/\Gamma=39$.  The intensities are $\Omega =39\Gamma$ (left), $\Omega =43\Gamma$ (middle), and $\Omega =47\Gamma$ (right), as in Fig.~1 of Ref.~\cite{Soding1997}.}
\end{figure}

While all the prior examples relied solely on spontaneous emission, {\tt pylcp} can also calculate forces due to stimulated emission.
A common setup for stimulated optical forces is the bi-chromatic force, which involves creating counterpropogating $\pi$ pulse trains that transfer the atom to the excited state from one direction and stimulate emission into the ground state from the opposite direction.
This stimulated optical force was examined theoretically and demonstrated experimentally in Ref.~\cite{Soding1997}.
In Ref.~\cite{Soding1997}, the pulses were made by shining two frequencies of light from both directions, each frequency detuned by $\pm\Delta$ from the excited state.
The phases between all four frequencies must be well established to ensure the proper direction of stimulated emission.
Fig.~\ref{fig:bichromatic} shows the calculation of the force for a two-level atom in this laser arrangement using {\tt pylcp}.
For each intensity, the force exceeds the spontaneous force limit of $\hbar k \Gamma$, and has a host of curious features including Doppleron resonances.
The bi-chromatic force simulations in {\tt pylcp} could easily be extended to more complicated systems such as real atoms or molecules, allowing examination of bi-chromatic force operation on non-cycling transitions.

\section{Conclusion}
We have presented a software package that integrates a set of lasers, a magnetic field, and an atomic or molecular Hamiltonian to accurately simulate laser cooling physics by automatically generating the optical Bloch equations or rate equations.
In this paper, we have demonstrated a host of useful computations previously documented in the literature, including Doppler and sub-Doppler cooling, capture of atoms and final equilibrium temperature of a MOT, limitations of trapping when additional level structure is added, and even recoil-limited MOTs such as $^1 \mbox{S}_{0}\rightarrow\, ^3\mbox{P}_1$ MOTs for Sr.
We also demonstrated phenomena that rely on time dependence of the applied fields, including adiabatic rapid passage and stimulated Raman adiabatic passage.
Finally, our examples also include forces that arise from coherent phenomena, like the bi-chromatic force.

We have begun to deploy {\tt pylcp} to understand the properties of grating MOTs~\cite{Lee2013a, Nshii2013, Barker2019, Sitaram2020} with complicated level structure.
These MOTs have quite peculiar properties, including a general spin polarization, asymmetric damping forces leading to asymmetric temperatures, and equilibrium trap positions that differ from the magnetic field zero.
Indeed, they also appear to be sensitive to the internal level structure of the atom.
For example, whereas a $^{88}$Sr grating MOT was realized recently on an $F=0\rightarrow F'=1$ transition, a $^{87}$Sr atom with $F=9/2\rightarrow F'= 11/2$ transition was not realized in the same apparatus~\cite{Sitaram2020}.
Studies are currently underway to determine whether a similar phenomenon as that in Fig.~\ref{fig:mot_F1_to_F2} are responsible for this failure in a grating MOT geometry.

While versatile, {\tt pylcp} has some limitations that may be alleviated in a future version.
First, our code does not include the possibility of static electric fields that can shift states within a manifold.
Second, our governing equations are only valid in the single-atom limit, where we neglect secondary photon scattering and attenuation of a laser due to absorption.
These effects are known to limit the achievable density and temperature in a MOT~\cite{Walker1990, Townsend1995, Camara2014}.
Thus, {\tt pylcp} is only valid in the low-density limit, which makes it less valuable in simulating magneto-optical traps and molasses that are typically used on quantum degenerate gas experiments.
Nevertheless, we anticipate that {\tt pylcp} is sufficiently functional to begin analyzing novel forms of laser cooling and designing efficient, miniaturized systems for laser cooling.





\bibliographystyle{elsarticle-num}
\bibliography{pylcp}

\begin{thebibliography}{10}
\expandafter\ifx\csname url\endcsname\relax
  \def\url#1{\texttt{#1}}\fi
\expandafter\ifx\csname urlprefix\endcsname\relax\def\urlprefix{URL }\fi
\expandafter\ifx\csname href\endcsname\relax
  \def\href#1#2{#2} \def\path#1{#1}\fi

\bibitem{Ludlow2015}
A.~D. Ludlow, M.~M. Boyd, J.~Ye, E.~Peik, P.~O. Schmidt, {Optical atomic
  clocks}, Reviews of Modern Physics 87~(2) (2015) 637.
\newblock \href {http://dx.doi.org/10.1103/RevModPhys.87.637}
  {\path{doi:10.1103/RevModPhys.87.637}}.

\bibitem{Campbell2017}
S.~L. Campbell, R.~B. Hutson, G.~E. Marti, A.~Goban, N.~{Darkwah Oppong}, R.~L.
  McNally, L.~Sonderhouse, J.~M. Robinson, W.~Zhang, B.~J. Bloom, J.~Ye, {A
  Fermi-degenerate three-dimensional optical lattice clock}, Science 358 (2017)
  90--94.
\newblock \href {http://dx.doi.org/10.1126/science.aam5538}
  {\path{doi:10.1126/science.aam5538}}.

\bibitem{Elvin2019}
R.~Elvin, G.~W. Hoth, M.~Wright, B.~Lewis, J.~P. McGilligan, A.~S. Arnold,
  P.~F. Griffin, E.~Riis, {Cold-atom clock based on a diffractive optic},
  Optics Express 27 (2019) 38359.
\newblock \href {http://dx.doi.org/10.1364/oe.378632}
  {\path{doi:10.1364/oe.378632}}.

\bibitem{Cronin2009}
A.~D. Cronin, J.~Schmiedmayer, D.~E. Pritchard, {Optics and interferometry with
  atoms and molecules}, Rev. Mod. Phys. 81~(3) (2009) 1051--1129.
\newblock \href {http://dx.doi.org/10.1103/RevModPhys.81.1051}
  {\path{doi:10.1103/RevModPhys.81.1051}}.

\bibitem{Becker2018}
D.~Becker, M.~D. Lachmann, S.~T. Seidel, H.~Ahlers, A.~N. Dinkelaker,
  J.~Grosse, O.~Hellmig, H.~M{\"{u}}ntinga, V.~Schkolnik, T.~Wendrich,
  A.~Wenzlawski, B.~Weps, R.~Corgier, T.~Franz, N.~Gaaloul, W.~Herr,
  D.~L{\"{u}}dtke, M.~Popp, S.~Amri, H.~Duncker, M.~Erbe, A.~Kohfeldt,
  A.~Kubelka-Lange, C.~Braxmaier, E.~Charron, W.~Ertmer, M.~Krutzik,
  C.~L{\"{a}}mmerzahl, A.~Peters, W.~P. Schleich, K.~Sengstock, R.~Walser,
  A.~Wicht, P.~Windpassinger, E.~M. Rasel, {Space-borne Bose–Einstein
  condensation for precision interferometry}, Nature 562~(7727) (2018)
  391--395.
\newblock \href {http://dx.doi.org/10.1038/s41586-018-0605-1}
  {\path{doi:10.1038/s41586-018-0605-1}}.

\bibitem{Xu2019}
V.~Xu, M.~Jaffe, C.~D. Panda, S.~L. Kristensen, L.~W. Clark, H.~M{\"{u}}ller,
  {Probing gravity by holding atoms for 20 seconds}, Science 366~(6466) (2019)
  745--749.
\newblock \href {http://dx.doi.org/10.1126/science.aay6428}
  {\path{doi:10.1126/science.aay6428}}.

\bibitem{Kitching2018}
J.~Kitching, {Chip-scale atomic devices}, Applied Physics Reviews 5 (2018)
  031302.
\newblock \href {http://dx.doi.org/10.1063/1.5026238}
  {\path{doi:10.1063/1.5026238}}.

\bibitem{Cohen2019}
Y.~Cohen, K.~Jadeja, S.~Sula, M.~Venturelli, C.~Deans, L.~Marmugi, F.~Renzoni,
  {A cold atom radio-frequency magnetometer}, Applied Physics Letters 114
  (2019) 073505.
\newblock \href {http://dx.doi.org/10.1063/1.5084004}
  {\path{doi:10.1063/1.5084004}}.

\bibitem{Pezze2018}
L.~Pezz{\`{e}}, A.~Smerzi, M.~K. Oberthaler, R.~Schmied, P.~Treutlein, {Quantum
  metrology with nonclassical states of atomic ensembles}, Reviews of Modern
  Physics 90 (2018) 035005.
\newblock \href {http://dx.doi.org/10.1103/RevModPhys.90.035005}
  {\path{doi:10.1103/RevModPhys.90.035005}}.

\bibitem{Bruzewicz2019}
C.~D. Bruzewicz, J.~Chiaverini, R.~McConnell, J.~M. Sage, {Trapped-ion quantum
  computing: Progress and challenges}, Applied Physics Reviews 6 (2019) 021314.
\newblock \href {http://dx.doi.org/10.1063/1.5088164}
  {\path{doi:10.1063/1.5088164}}.

\bibitem{Safronova2018}
M.~S. Safronova, D.~Budker, D.~Demille, D.~F.~J. Kimball, A.~Derevianko, C.~W.
  Clark, \href{https://doi.org/10.1103/RevModPhys.90.025008}{{Search for new
  physics with atoms and molecules}}, Reviews of Modern Physics 90 (2018)
  025008.
\newblock \href {http://dx.doi.org/10.1103/RevModPhys.90.025008}
  {\path{doi:10.1103/RevModPhys.90.025008}}.
\newline\urlprefix\url{https://doi.org/10.1103/RevModPhys.90.025008}

\bibitem{hutzler2020}
N.~R. Hutzler, Polyatomic molecules as quantum sensors for fundamental physics
  (2020).
\newblock \href {http://arxiv.org/abs/2008.03398} {\path{arXiv:2008.03398}}.

\bibitem{Norrgard2019}
E.~B. Norrgard, D.~S. Barker, S.~Eckel, J.~A. Fedchak, N.~N. Klimov,
  J.~Scherschligt,
  \href{http://dx.doi.org/10.1038/s42005-019-0181-1}{{Nuclear-spin dependent
  parity violation in optically trapped polyatomic molecules}}, Communications
  Physics 2 (2019) 77.
\newblock \href {http://dx.doi.org/10.1038/s42005-019-0181-1}
  {\path{doi:10.1038/s42005-019-0181-1}}.
\newline\urlprefix\url{http://dx.doi.org/10.1038/s42005-019-0181-1}

\bibitem{Foot2005}
C.~Foot, \href{https://books.google.com/books?id=\_CoSDAAAQBAJ}{{Atomic
  Physics}}, Oxford Master Series in Physics, Oxford University Press Inc., New
  York, 2005.
\newline\urlprefix\url{https://books.google.com/books?id=\_CoSDAAAQBAJ}

\bibitem{Metcalf1999}
H.~Metcalf, P.~van~der Straten,
  \href{https://books.google.com/books?id=RJXwBwAAQBAJ}{{Laser Cooling and
  Trapping}}, Graduate Texts in Contemporary Physics, Springer-Verlag New York
  Inc., New York, 1999.
\newline\urlprefix\url{https://books.google.com/books?id=RJXwBwAAQBAJ}

\bibitem{Lett1989}
P.~D. Lett, W.~D. Phillips, S.~L. Rolston, C.~E. Tanner, R.~N. Watts, C.~I.
  Westbrook, {Optical molasses}, Journal of the Optical Society of America B
  6~(11) (1989) 2084.
\newblock \href {http://dx.doi.org/10.1364/JOSAB.6.002084}
  {\path{doi:10.1364/JOSAB.6.002084}}.

\bibitem{Dalibard1989}
J.~Dalibard, C.~Cohen-Tannoudji, {Laser cooling below the Doppler limit by
  polarization gradients: simple theoretical models}, Journal of the Optical
  Society of America B 6~(11) (1989) 2023.
\newblock \href {http://dx.doi.org/10.1364/JOSAB.6.002023}
  {\path{doi:10.1364/JOSAB.6.002023}}.

\bibitem{Rushton2014}
J.~A. Rushton, M.~Aldous, M.~D. Himsworth, {Contributed Review: The feasibility
  of a fully miniaturized magneto-optical trap for portable ultracold quantum
  technology}, Rev. Sci. Instruments 85~(12).
\newblock \href {http://dx.doi.org/10.1063/1.4904066}
  {\path{doi:10.1063/1.4904066}}.

\bibitem{Bartolotta2018}
J.~P. Bartolotta, M.~A. Norcia, J.~R.~K. Cline, J.~K. Thompson, M.~J. Holland,
  {Laser cooling by sawtooth-wave adiabatic passage}, Phys. Rev. A 98 (2018)
  023404.
\newblock \href {http://dx.doi.org/10.1103/PhysRevA.98.023404}
  {\path{doi:10.1103/PhysRevA.98.023404}}.

\bibitem{Raab1987}
E.~L. Raab, M.~Prentiss, A.~Cable, S.~Chu, D.~E. Pritchard, {Trapping of
  Neutral Sodium Atoms with Radiation Pressure}, Physical Review Letters
  59~(23) (1987) 2631.
\newblock \href {http://dx.doi.org/10.1103/PhysRevLett.59.2631}
  {\path{doi:10.1103/PhysRevLett.59.2631}}.

\bibitem{Johansson2012}
J.~R. Johansson, P.~D. Nation, F.~Nori, {QuTiP: An open-source Python framework
  for the dynamics of open quantum systems}, Computer Physics Communications
  183 (2012) 1760.
\newblock \href {http://dx.doi.org/10.1016/j.cpc.2012.02.021}
  {\path{doi:10.1016/j.cpc.2012.02.021}}.

\bibitem{Johansson2013}
J.~R. Johansson, P.~D. Nation, F.~Nori, {QuTiP 2: A Python framework for the
  dynamics of open quantum systems}, Computer Physics Communications 184 (2013)
  1234.
\newblock \href {http://dx.doi.org/10.1016/j.cpc.2012.11.019}
  {\path{doi:10.1016/j.cpc.2012.11.019}}.

\bibitem{ADM}
\url{http://rochesterscientific.com/ADM/}.

\bibitem{pylcp}
\url{https://www.github.com/jqiAMO/pylcp}.

\bibitem{pylcpdocs}
\url{https://python-laser-cooling-physics.readthedocs.io/en/stable}.

\bibitem{Gordon1980}
J.~P. Gordon, A.~Ashkin, {Motion of atoms in a radiation trap}, Physical Review
  A 21~(5) (1980) 1606--1617.
\newblock \href {http://dx.doi.org/10.1103/PhysRevA.21.1606}
  {\path{doi:10.1103/PhysRevA.21.1606}}.

\bibitem{Ungar1989}
P.~J. Ungar, D.~S. Weiss, E.~Riis, S.~Chu, {Optical molasses and multilevel
  atoms: theory}, Journal of the Optical Society of America B 6~(11) (1989)
  2058.
\newblock \href {http://dx.doi.org/10.1364/JOSAB.6.002058}
  {\path{doi:10.1364/JOSAB.6.002058}}.

\bibitem{Devlin2018}
J.~A. Devlin, M.~R. Tarbutt, {Laser cooling and magneto-optical trapping of
  molecules analyzed using optical Bloch equations and the
  Fokker-Planck-Kramers equation}, Physical Review A 98~(6) (2018) 1--14.
\newblock \href {http://dx.doi.org/10.1103/PhysRevA.98.063415}
  {\path{doi:10.1103/PhysRevA.98.063415}}.

\bibitem{Dalibard1982}
J.~Dalibard, J.~Dupont-Roc, C.~Cohen-Tannoudji, {Vacuum Fluctuations and
  Radiation Reaction - Identification of Their Respective Contributions.},
  Journal de physique Paris 43~(11) (1982) 1617--1638.
\newblock \href {http://dx.doi.org/10.1051/jphys:0198200430110161700}
  {\path{doi:10.1051/jphys:0198200430110161700}}.

\bibitem{Milonni1984}
P.~W. Milonni, {Why spontaneous emission?}, American Journal of Physics 52~(4)
  (1984) 340--343.
\newblock \href {http://dx.doi.org/10.1119/1.13886}
  {\path{doi:10.1119/1.13886}}.

\bibitem{Tarbutt2015}
M.~R. Tarbutt, {Magneto-optical trapping forces for atoms and molecules with
  complex level structures}, New Journal of Physics 17~(1) (2015) 015007.
\newblock \href {http://dx.doi.org/10.1088/1367-2630/17/1/015007}
  {\path{doi:10.1088/1367-2630/17/1/015007}}.

\bibitem{Lu2005}
T.~Lu, X.~Miao, H.~Metcalf, {Bloch theorem on the Bloch sphere}, Phys. Rev. A
  71 (2005) 061405.
\newblock \href {http://dx.doi.org/10.1103/PhysRevA.71.061405}
  {\path{doi:10.1103/PhysRevA.71.061405}}.

\bibitem{Boller1991}
K.-J. Boller, A.~Imamoglu, S.~E. Harris, {Observation of Electromagnetically
  Induced Transparency}, Phys. Rev. Lett. 66~(20) (1991) 2593.
\newblock \href {http://dx.doi.org/10.1103/PhysRevLett.66.2593}
  {\path{doi:10.1103/PhysRevLett.66.2593}}.

\bibitem{Bergmann1998}
K.~Bergmann, H.~Theuer, B.~W. Shore, Coherent population transfer among quantum
  states of atoms and molecules, Rev. Mod. Phys. 70 (1998) 1003.
\newblock \href {http://dx.doi.org/10.1103/RevModPhys.70.1003}
  {\path{doi:10.1103/RevModPhys.70.1003}}.

\bibitem{Chu1985}
S.~Chu, L.~Hollberg, J.~E. Bjorkholm, A.~Cable, A.~Ashkin, {Three-Dimensional
  Viscous Confinement and Cooling of Atoms by Resonance Radiation Pressure},
  Physical Review Letters 55~(1) (1985) 48.

\bibitem{Devlin2016}
J.~A. Devlin, M.~R. Tarbutt, {Three-dimensional Doppler, polarization-gradient,
  and magneto-optical forces for atoms and molecules with dark states}, New
  Journal of Physics 18~(12) (2016) 123017.
\newblock \href {http://dx.doi.org/10.1088/1367-2630/18/12/123017}
  {\path{doi:10.1088/1367-2630/18/12/123017}}.

\bibitem{Grier2013}
A.~T. Grier, I.~Ferrier-Barbut, B.~S. Rem, M.~Delehaye, L.~Khaykovich,
  F.~Chevy, C.~Salomon, {$\Lambda$-enhanced sub-Doppler cooling of lithium
  atoms in D$_1$ gray molasses}, Physical Review A 87~(6) (2013) 063411.
\newblock \href {http://dx.doi.org/10.1103/PhysRevA.87.063411}
  {\path{doi:10.1103/PhysRevA.87.063411}}.

\bibitem{Haubrich1993}
D.~Haubrich, A.~H{\"{o}}pe, D.~Meschede, {A simple model for optical capture of
  atoms in strong magnetic quadrupole fields}, Optics Communications 102~(3-4)
  (1993) 225--230.
\newblock \href {http://dx.doi.org/10.1016/0030-4018(93)90387-K}
  {\path{doi:10.1016/0030-4018(93)90387-K}}.

\bibitem{Mukaiyama2003}
T.~Mukaiyama, H.~Katori, T.~Ido, Y.~Li, M.~Kuwata-Gonokami, {Recoil-limited
  laser cooling of $^{87}$Sr atoms near the Fermi temperature}, Phys. Rev.
  Lett. 90~(11) (2003) 113002.
\newblock \href {http://dx.doi.org/10.1103/PhysRevLett.90.113002}
  {\path{doi:10.1103/PhysRevLett.90.113002}}.

\bibitem{Hanley2018}
R.~K. Hanley, P.~Huillery, N.~C. Keegan, A.~D. Bounds, D.~Boddy, R.~Faoro,
  M.~P. Jones, {Quantitative simulation of a magneto-optical trap operating
  near the photon recoil limit}, Journal of Modern Optics 65~(5-6) (2018)
  667--676.
\newblock \href {http://dx.doi.org/10.1080/09500340.2017.1401679}
  {\path{doi:10.1080/09500340.2017.1401679}}.

\bibitem{Katori1999}
H.~Katori, T.~Ido, Y.~Isoya, M.~Kuwata-Gonokami,
  \href{http://prl.aps.org/abstract/PRL/v82/i6/p1116{\_}1}{{Magneto-Optical
  Trapping and Cooling of Strontium Atoms down to the Photon Recoil
  Temperature}} 82~(6) (1999) 1116.
\newblock \href {http://dx.doi.org/10.1103/PhysRevLett.82.1116}
  {\path{doi:10.1103/PhysRevLett.82.1116}}.
\newline\urlprefix\url{http://prl.aps.org/abstract/PRL/v82/i6/p1116{\_}1}

\bibitem{Loftus2004b}
T.~H. Loftus, T.~Ido, A.~D. Ludlow, M.~M. Boyd, J.~Ye, {Narrow Line Cooling:
  Finite Photon Recoil Dynamics}, Physical Review Letters 93 (2004) 073003.
\newblock \href {http://dx.doi.org/10.1103/PhysRevLett.93.073003}
  {\path{doi:10.1103/PhysRevLett.93.073003}}.

\bibitem{Stellmer2013}
S.~Stellmer, {Degenerate quantum gases of strontium}, Ph.D. thesis, University
  of Innsbruck (2013).

\bibitem{Flemming1997}
J.~Flemming, A.~M. Tuboy, D.~M. B.~P. Milori, L.~G. Marcassa, S.~C. Zilio,
  V.~S. Bagnato, {Magneto-optical trap for sodium atoms operating on the D1
  line}, Optics Communications 135 (1997) 269.
\newblock \href {http://dx.doi.org/10.1016/S0030-4018(96)00660-8}
  {\path{doi:10.1016/S0030-4018(96)00660-8}}.

\bibitem{Tarbutt2015a}
M.~R. Tarbutt, T.~C. Steimle, {Modeling magneto-optical trapping of CaF
  molecules}, Physical Review A 92~(5) (2015) 053401.
\newblock \href {http://dx.doi.org/10.1103/PhysRevA.92.053401}
  {\path{doi:10.1103/PhysRevA.92.053401}}.

\bibitem{Soding1997}
J.~S{\"{o}}ding, R.~Grimm, Y.~Ovchinnikov, P.~Bouyer, C.~Salomon,
  {Short-Distance Atomic Beam Deceleration with a Stimulated Light Force},
  Physical Review Letters 78 (1997) 1420--1423.
\newblock \href {http://dx.doi.org/10.1103/PhysRevLett.78.1420}
  {\path{doi:10.1103/PhysRevLett.78.1420}}.

\bibitem{Lee2013a}
J.~Lee, J.~A. Grover, L.~A. Orozco, S.~L. Rolston, {Sub-Doppler cooling of
  neutral atoms in a grating magneto-optical trap}, Journal of the Optical
  Society of America B 30~(11) (2013) 2869.
\newblock \href {http://dx.doi.org/10.1364/JOSAB.30.002869}
  {\path{doi:10.1364/JOSAB.30.002869}}.

\bibitem{Nshii2013}
C.~C. Nshii, M.~Vangeleyn, J.~P. Cotter, P.~F. Griffin, E.~A. Hinds, C.~N.
  Ironside, P.~See, a.~G. Sinclair, E.~Riis, a.~S. Arnold, {A surface-patterned
  chip as a strong source of ultracold atoms for quantum technologies}, Nature
  Nanotechnology 8~(5) (2013) 321--324.
\newblock \href {http://dx.doi.org/10.1038/nnano.2013.47}
  {\path{doi:10.1038/nnano.2013.47}}.

\bibitem{Barker2019}
D.~S. Barker, E.~B. Norrgard, N.~N. Klimov, J.~A. Fedchak, J.~Scherschligt,
  S.~Eckel, {Single-Beam Zeeman Slower and Magneto-Optical Trap Using a
  Nanofabricated Grating}, Physical Review Applied 11 (2019) 064023.
\newblock \href {http://dx.doi.org/10.1103/PhysRevApplied.11.064023}
  {\path{doi:10.1103/PhysRevApplied.11.064023}}.

\bibitem{Sitaram2020}
A.~Sitaram, P.~K. Elgee, G.~K. Campbell, N.~N. Klimov, S.~Eckel, D.~S. Barker,
  {Confinement of an alkaline-earth element in a grating magneto-optical trap},
  Review of Scientific Instruments 91 (2020) 103202.
\newblock \href {http://dx.doi.org/10.1063/5.0019551}
  {\path{doi:10.1063/5.0019551}}.

\bibitem{Walker1990}
T.~Walker, D.~Sesko, C.~Wieman, Collective behavior of optically trapped
  neutral atoms, Phys. Rev. Lett. 64 (1990) 408--411.
\newblock \href {http://dx.doi.org/10.1103/PhysRevLett.64.408}
  {\path{doi:10.1103/PhysRevLett.64.408}}.

\bibitem{Townsend1995}
C.~G. Townsend, N.~H. Edwards, C.~J. Cooper, K.~P. Zetie, C.~J. Foot, A.~M.
  Steane, P.~Szriftgiser, H.~Perrin, J.~Dalibard, {Phase-space density in the
  magneto-optical trap}, Phys. Rev. A 52~(2) (1995) 1423--1440.
\newblock \href {http://dx.doi.org/10.1103/PhysRevA.52.1423}
  {\path{doi:10.1103/PhysRevA.52.1423}}.

\bibitem{Camara2014}
A.~Camara, R.~Kaiser, G.~Labeyrie, {Scaling behavior of a very large
  magneto-optical trap}, Phys. Rev. A 90~(6) (2014) 063404.
\newblock \href {http://arxiv.org/abs/1410.1832} {\path{arXiv:1410.1832}},
  \href {http://dx.doi.org/10.1103/PhysRevA.90.063404}
  {\path{doi:10.1103/PhysRevA.90.063404}}.

\end{thebibliography}

\end{document}